 \documentclass[10pt,preprint]{aastex63}

 \usepackage{natbib}
\def\a{\"a}
\def\o{\"o}
\def\u{\"u}
\def\A{\"A}
\def\O{\"O}
\def\U{\"U}

\def\ang{\AA}

\def\gapprox{\lower.4ex\hbox{$\;\buildrel >\over{\scriptstyle\sim}\;$}}
\def\lapprox{\lower.4ex\hbox{$\;\buildrel <\over{\scriptstyle\sim}\;$}}
\def\captio#1{\caption{\small {#1} \normalsize}}
\shortauthors{Aschwanden}
\shorttitle{SOC}

\begin{document}

\renewcommand{\topfraction}{0.95}
\renewcommand{\bottomfraction}{0.95}
\renewcommand{\textfraction}{0.05}
\renewcommand{\floatpagefraction}{0.95}
\renewcommand{\dbltopfraction}{0.95}
\renewcommand{\dblfloatpagefraction}{0.95}

\title{New Trends in Astrophysical Self-Organized Criticality}
 
\author{Markus J. Aschwanden}
\affil{Lockheed Martin, Solar and Astrophysics Laboratory (LMSAL),
       Advanced Technology Center (ATC),
       A021S, Bldg.252, 3251 Hanover St.,
       Palo Alto, CA 94304, USA;
       e-mail: markus.josef.aschwanden@gmail.com}

\begin{abstract}
This review is focused on recent {\sl Self-Organized Criticality
(SOC)} literature of astrophysical phenomena, covering 
the last decade of (2015-2025), while previous SOC literature 
(1987-2014) is reviewed elsewhere. The selection of literature
is mostly based on searches with the NASA-supported 
{\sl Astrophysics Data System (ADS)}. The discussed 
astrophysical SOC phenomena are subdivided into 
solar flares, 
solar atmosphere (photosphere, chromosphere, corona), 
heliospheric systems (coronal mass ejections, 
solar wind, solar energetic particles), 
planetary systems (asteroids and small bodies, 
lunar cratering, Saturnian ring systems, 
magnetospheric systems), 
stellar flares, and
galactic systems (pulsar glitches, gamma ray bursts, 
soft gamma-ray repeaters, supergiant fast X-ray transients,
fast transient radio bursts, magnetars, blazars, black holes).
\end{abstract}
 
\clearpage

\section{Introduction}

Four reviews were featured in the journal 
{\sl Space Science Reviews (SSRv)}, Volume 198, 
which are dedicated to the Special Issue of {\sl 25 
Years of Self-Organized Criticality}, covering 
concepts and controversies (Watkins et al.~2016),
solar and astrophysics (Aschwanden et al.~2016b),
space and laboratory plasmas (Sharma et al.~2016), and
numerical detection methods (McAteer et al.~2016).
Furthermore, textbooks are available 
(Aschwanden 2011, 2025a; Pruessner 2012).

Watkins et al.~(2016) stated that {\sl SOC has been
one of the most stimulating concepts to come out of
statistical mechanics and condensed matter theory
in the last few decades, and it has played a significant
role in the development of complexity science}.
At the same time it was admitted that SOC developed
into a direction that is different from the original
creators. {\sl SOC phenomena continue to live a
number of parallel lives in various fields}, such as
astrophysics, solar physics, magnetospheric physics, 
statistical mechanics, material sciences,
condensed matter theory, laboratory fusion plasmas, 
seismology, geophysics, biological evolution,
neuroscience, sociology, ecology, epidemics and pandemics
(including the 5-year era 2020-2025 of Covid-19), 
to name a few. Solar physics and stellar physics 
are considered as subfields of astrophysics. 
Watkins et al.~(2016) criticize the generalization 
of SOC concepts and point out that {\sl one source
of erroneous beliefs (sometimes found in the
literature) is that everything that is avalanching
must be critical and self-organized, or, conversely,
that everything that displays long-ranged correlations
or a power law must be an instance of (self-organized)
criticality,} which, however, are necessary but insufficient 
conditions. We reconcile this problem by accepting
multiple SOC definitions, such as the cellular automaton 
approach of the Bak-Tang-Wiesenfeld (BTW) model on one side
(Bak et al.~1987), 
and macroscopic physical scaling laws with power law-like 
size distributions
on the other side. One might consider this development
as a paradigm shift or a generalization of SOC-like models. 

The second of the four SSRv reviews 
(Aschwanden et al.~2016b) presents an account
of SOC phenomena in astrophysics at large, which includes 
solar physics, stellar physics, planetary physics, 
and galactic physics. The parameters of
interest are the size distribution functions (of fluxes,
fluences, time durations) and the waiting time distribution 
functions. The theoretical interpretation of SOC phenomena
is primarily based on physical scaling laws, rather than
cellular automaton mechanisms, which are extensively
reviewed in the textbook of Pruessner (2012). 

The third of the four SSRv reviews 
(Sharma et al.~2016) expands the SOC applications to 
space plasmas (heliosphere, magnetosphere, solar wind,
space weather, geomagnetic storms, substorms) and
laboratory plasmas. Many plasmas are unstable,
nonlinear, dissipative, exhibit multi-scale behavior,
non-equilibrium systems, intermittency, and turbulence.  
Laboratory plasma physics deals with tokamaks, 
stellarators, plasma fusion conditions, and confined plasmas.

The fourth of the four SSRv reviews (McAteer et al.~2016)
discusses methods of detecting SOC, from correlations
to complexity and critical quantities. In particular,
the auto-correlation and the structure function are
described in more detail. Emphasis is given to
feature detection in the spatial domain, in the
temporal domain, and in the spatio-temporal domain,
which are precursors of the {\sl Artificial 
Intelligence (AI)} efforts, pursued nowadays. 

A fifth review of SOC phenomena in astrophysics
is presented in the recent textbook {\sl ``Power
laws in astrophysics. Self-organized criticality
systems''} (Aschwanden 2025a), which covers the entire
SOC era of 1987-2025. This book is organized into
two parts, a theoretical part and an observational
part. The theoretical part covers the fundamentals,
the power law size distributions, and the waiting 
time distributions, which are needed in order to
understand the SOC-related observations.
The observational part covers SOC measurements
of astrophysical phenomena from solar flares to
galactic systems. What is new in our review here
are updates and new trends of SOC modeling during
the last decade (2015-2025).


\section{Solar Flares}

\subsection{Size Distributions}

Tests of SOC models require observational
measurements of fundamental physical parameters, 
which have been determined in $\approx 400$ M- 
and X-class flares, using SDO/AIA images in 
EUV wavelengths (Aschwanden et al.~2015a). 
The corresponding size distributions are shown in Fig.~(1).
Some of these fundamental parameters are given 
(with approximative ranges): 
the length scale of flare areas $(L=10^{8.2}-10^{9.7}$ cm);
the peak electron temperature $(T_e=10^{5.7}-10^{7.4}$ K);
the peak electron density $(n_e=10^{10.3}-10^{11.8}$ cm$^{-3})$; 
and the (multi-thermal) energy ($E_{th}=10^{26.8}-10^{32.0}$ erg). 
Obviously, the electron temperature does not qualify as a valid 
SOC parameter, based on the power law with a positive slope. 
The size distributions of solar flares within these
parameter ranges exhibit power law slopes with values of
$\alpha_L=3.3\pm0.3$ for length scales, 
$\alpha_V=1.7\pm0.2$ for flare volumina, and 
$\alpha_{th}=1.8\pm0.2$ for thermal energies
(Aschwanden et al.~2015a), which are consistent 
with the theoretical predictions of the 
fractal-diffusive self-organized criticality (FD-SOC) 
model (Table 1). 

Some generalization of SOC models can be obtained by
defining energies in terms of area-like 2-D geometries, 
rather than volume-like 3-D geometries, which can 
reconcile the difference of power law slopes in flares 
and nanoflares. The flare volume is defined then as
$V = A h = L^2$ in the 2-D world, with $h$ being a 
constant, and is defined as $V = A^{3/2} = L^3$ in 
the 3-D world, with $h=A^{1/2} = L$ being the 
extension along the line-of-sight, in the isotropic 
assumption (Aschwanden 2022a). 

The accuracy of power law distribution functions
with obvious deviations from ideal power laws
has been explained in terms of various side effects:
(i) a physical threshold of an instability
(Aschwanden 2015);
ii) incomplete sampling of the smallest events
due to limited instrumental sensitivity;
(iii) contamination by an event-unrelated background
(Li et al.~2016; 
(iv) truncation of the largest events due to
finite system size effects (Aschwanden 2021). 
These effects have been modeled in terms of
Pareto-type II, or Lomax distribution function,
\begin{equation}
	N(x) dx = n_0 (x_0+x)^{-\alpha_x} \ .
\end{equation}
where $x_0$ is a threshold parameter, and
$x_0 \mapsto 0$ converges to an ideal power law 
function. Finite system-size effects occur 
at the largest sizes of active regions, at length
scales of $L \lapprox 200$ Mm in the case of
solar flares (Aschwanden 2021). Additional 
effects may involve: 
(v) The time variability of the SOC driver, 
(vi) the violation of time scale separation 
(when the waiting time is not much larger 
than the flare duration), 
(vii) a fast driver with a non-stationary flare rate 
(Aschwanden 2019c), 
and (ix) the choice of the waiting time 
distribution function. 
(x) Strong deviations from power laws can arise 
also from EUV filter degradation (Ryan et al.~2016). 
(xi) Fitting a log-normal instead
of a power law yielded better results in some
cases, but is very sensitive to the 
(event-unrelated) background subtraction
(Verbeeck et al.~2019).
Examples of a power law-like size distributions are
given in Fig.~(2), based on the Kepler stellar flare 
catalog, which illustrates the background-subtracted 
fitting with a Pareto Type II function (PM),
and finite system size effects (FM).  


\subsection{Waiting Time Distributions (WTD)}

The waiting time $T$ is defined by the time interval
between two subsequent events in a time-sorted event list, 
$T_i=t_{i+1}-t_i$ for
$i=1,...,n$. If statistical events are produced by
a linear random process, the resulting {\sl waiting time 
distribution (WTD)} function follows Poisson statistics, 
which falls off approximately as an exponential 
distribution function,
\begin{equation}
	N(T) dT = \lambda_0 \exp[-\lambda_0 T] dT ,
\end{equation}
where $\lambda_0$ is the mean event rate (or flare
rate in the case of solar flares). Hence, it appears
that we can straightforwardly identify whether the observed 
size distribution function falls off like an exponential 
function, or if it falls off like a power law distribution
function, which allows to test the interpretation of a 
{\sl self-organized criticality (SOC)} process. 
Since most of observed size distributions
of SOC phenomena exhibit a power law size distribution, 
it was concluded that these data do not follow the
expected Poisson statistics, and thus cannot be
SOC processes. However, this logic is flawed because
it assumes a stationary flare rate $\lambda_0$, while
in reality the flare rate $\lambda(t)$ is always 
non-stationary and thus time-dependent. For instance, the
flaring rate varies by a factor of 
$(\lambda_{\rm max}/\lambda_{\rm min}) \approx 10^2-10^3$ 
during a solar cycle, in the case of solar flares.
A realistic non-stationary WTD can be expressed by
a convolution of exponentially-falling size distributions,
\begin{equation}
	N(T_i) dT = {\sum_i \lambda_i^2 T_i 
	\exp[-\lambda_i T_i] \over \sum_i \lambda_i T_i } \ .
\end{equation}
In modeling of the WTD function,
the variability of the nonstationary flaring rate 
requires a multi-component (Bayesian block) decomposition,
where each component is quasi-stationary and has a
Poissonian origin (Aschwanden et al.~2021).
The WTD functions of fast-driven and slowly-driven
SOC models, stationary and nonstationary SOC models,
are depicted in Fig.~(3), (Aschwanden 2019c).
Along the same line of reasoning, a correlation was 
found for the sunspot number and the WTD of solar flares, 
which is a proxy of the mean flare rate.
Parameterizing the time dependence of the flaring rate
with a polynomial function in the exponentially-growth phase,
\begin{equation}
	\lambda(t)=\lambda_0 \left( 
	{t \over T/2} \right)^p \ ,
\end{equation}
solar data exhibit a power law slope of $\alpha_{\rm WTD}=2.1-2.4$,
which agrees well with the theoretial prodiction
$\alpha_{\rm WTD}=2.0 + {1/p}$, where $p=1$ corresponds to
linear time evolutions, and $p=2$ corresponds to
nonlinear (approximately quadratic or 
exponentially-growing) evolution 
(Aschwanden and Johnson 2021; Aschwanden 2025a).

The lack of a correlation
between the waiting time and SOC avalanche
flux (amplitude) does not support SOC models
with long-term energy storage (Aschwanden 2019a), 
although attempts with small-number statistics
have been made (Hudson 2020).

The WTD statistics 
observed in a flare-productive active region
suggests that an individual active region
maintains some memory, in contrast to pure random
(uncorrelated) time structures (Lei et al.~2020).
Analyzing the rise times of time structures over 
a large time range, it was found that solar 
phenomena exhibit memory from hours to decades, 
where the longest memory is produced by the 
(magnetic) solar dynamo (Aschwanden and Johnson (2021).


\subsection{Cellular Automaton Models}

The {\sl Bak-Tang-Wiesenfeld (BTW)} model is the 
prototype of a {\sl cellular automaton (CA)} 
model for SOC processes, but all redistribution 
rules in existing CA models are {\sl ad hoc}. 
CA models come in many flavors, such as
isotropic or anisotropic, deterministic
or probabilistic, short-distance or long-distance
interacting, conservative or non-conservative
models (Farhang et al.~2019). Many models can
reproduce the observed power law size distributions,
and thus are not unique. A new CA model in which the
maximum possible amount of energy is released
at each redistribution has been constructed 
(Farhang et al.~2018; Lamarre et al.~2024), 
which is a conservative anisotropic CA model. 
Non-conservative deterministic models produce
the longest time window of event prediction,
while the stochastically-driven conservative
models perform less satisfactory (Thibeault et al.~2022).
CA models with two coupled lattices were constructed
to mimic ``sympathetic flaring'', but the authors
conclude that magnetic connectivity between
distinct active regions drives sympathetic flaring
on the Sun only weakly (Guite et al.~2025a), 
while observations with AIA/SDO, RHESSI and STIX 
suggest a sympathetic flaring rate of the order 
of 5\% (Guite at al.~2025b). 


\subsection{The Fractal-Diffusive SOC Model}

The {\sl fractal-diffusive Self-Organized Criticality (FD-SOC)}
model (Aschwanden 2014, 2015, 2022b, 2025b)
is based on four fundamental assumptions:
(i) The fractality,
(ii) the scale-freeness,
(iii) the flux-volume proportionality of incoherent processes, and
(iv) classical diffusion. 
Based on these four assumptions, the FD-SOC model predicts 
power law functions for the size distributions of SOC parameters,
as well as the power law slopes for each distribution.
The FD-SOC model makes quantitative predictions as a function of
the Euclidean space dimension $d$, i.e., 
$d=1$ for curvi-linear structures, 
$d=2$ for area-like geometries, and
$d=3$ for voluminous structures.
In the following we consider mostly the case for $d=3$. 

Instead of using the next-neighbor interactions of cellular
automata, as defined in the original SOC model of Bak et al.~(1987),
we use some universal scaling laws between various SOC parameters.
The spatial inhomogeneity of a SOC avalanche is expressed
in terms of the fractal dimension $D_d$.
Each fractal domain has a maximum fractal dimension of $D_d=d$,
a minimum value of $D_d=(d-1)$, which implies a mean value of 
$D_V=d-1/2=2.5$,
\begin{equation}
        D_V={(D_{\rm V,max} + D_{\rm V,min}) \over 2} = d-{1 \over 2}  \ .
\end{equation}
In the following we denote $D_3$ also as $D_V$. 
For most applications in the observed 3-D world, the dimensional
domain $d=3$ is appropriate, which implies a fractal dimension of
$D_V=2.5$. The fractal volume $V$ is then defined by the standard
(Hausdorff) fractal dimension $D_V$ in 3-D and the length scale
$L$ (Mandelbrot 1977),
\begin{equation}
        V(L) \propto L^{D_V} \ .
\end{equation}
The flux is assumed to be proportional to the avalanche volume
for the case of incoherent growth ($\gamma=1$),
but can be generalized for coherent growth ($\gamma \gapprox 2$),
\begin{equation}
        F \propto V^\gamma = \left( L^{D_V} \right)^\gamma \ .
\end{equation}
The spatio-temporal evolution is approximated with the assumption
of (classical) diffusive transport,
\begin{equation}
        L \propto T^{\beta/2} = T^{1/2} \ ,
\end{equation}
with the diffusion coefficient $\beta=1$.
Diffusive SOC models like an ordinary Brownian
random walk is also called {\sl invasion percolation}
in the literature, obeying $<x^2> \propto t$ 
(Watkins et al.~2016, p.20, footnote $^{13}$)

The statistics of SOC avalanches is quantified in terms of size
distributions (or occurrence frequency
distributions) that obey the scale-free probability distribution
function (Aschwanden 2014, 2015, 2022b, 2025b), expressed with the 
power law function, akas Zipf's law (Newman 2005),
\begin{equation}
        N(L)\ dL \propto L^{-d} dL \ .
\end{equation}
From the scale-free relationship, the power law slopes $\alpha_s$
of other SOC size parameters $s=[A,V,F,E,T]$ can be derived,
such as for the area $A$, the volume $V$, the flux $F$,
the fluence or energy $E$, and the duration $T$. The resulting
power law slopes $\alpha_s$ can then be obtained mathematically
by the method of variable substitution $s(L)$, by inserting the
inverse function $L(s)$ and its derivative $|dL/ds|$,
\begin{equation}
        N(s) ds = N[L(s)] \left| {dL \over ds} \right| dL
        = \ s^{-\alpha_s} ds \ ,
\end{equation}
such as for the flux $s=F$,
\begin{equation}
        \alpha_F = 1 + {(d-1) \over D_V \gamma} = {9 \over 5} = 1.80 \ ,
\end{equation}
for the fluence of energy $s=E$,
\begin{equation}
        \alpha_E = 1 + {(d-1) \over d \gamma} = {5 \over 3} \approx 1.67 \ .
\end{equation}
and for the time duration $s=T$,
\begin{equation}
        \alpha_T = 1 + (d-1)\beta/2 = 2 \ .
\end{equation}
Note that the flux has a fractal volume ($D_V$, Eq.~11), while the 
time-integrated fluence (or energy) has an Euclidean volume ($d$, Eq.~12). 
We call this model the standard FD-SOC model,
defined by [$d=3, \gamma=1, \beta=1$],
while the generalized FD-SOC model allows for variable coefficients
[$d, \gamma, \beta$] and alternative dimensionalities ($d = 1, 2$).
Some predictions of the FD-SOC model are juxtaposed to the observed
values in Table 1, which show satisfactory agreement.


\subsection{Nanoflares}

The largest events in a SOC size distribution are of highest
interest because of their potential for catastrophic predictions,
but the smallest events are equally interesting because of 
the diagnostics of their convergence or divergence of  
energy at the smallest events. The critical 
parameter is the power law slope of $\alpha_E=2.0$. 
If $\alpha_E<2$, most of the flare energies is dissipated
in the largest flares, while for $\alpha_E>2$, most of the 
energy is dissipated in the nanoflare regime (Hudson 1978). 
This information
is crucial for the identification of coronal heating processes.
According to the prediction of the FD-SOC model (Table 1), 
the power law
slopes of flux and energies are below the critical value of
$\alpha_E=2$, and thus one would expect that the largest
flares contribute most of the dissipated energy budget (such as 
in GOES M- and X-class flares), while 
(GOES B- and C-class) nanoflares contribute only
a negligible part. A reconciliation in terms of a two-component
hybrid size distribution has been proposed by 
Kawai and Imada (2022), where $\alpha_E > 2$ includes the
granule-scale/Parker's model, and $\alpha_E < 2$ entails the
sunspot-scale/CSHKP model. This line of reasoning is still quite
controversial, as the interpretation of the most recent
studies demonstrates.

There is a rapidly-growing nomenclature of nanoflare-like 
events, such as 
reconnection nanojets (Antolin et al.~2021), 
weak-emission events (Ulyanov et al.~2019),
``campfires'' (Alipour et al.~2022), 
small-scale heating events (Guerreiro et al.~2017), 
coronal hole and Quiet-Sun EUV intensity fluctuations 
(Cadavid et al.~2019), 
dynamic flaring non-potential structures
in supergranule networks (Chesny et al.~2016),
micro-sigmoids (Chesny et al.~2016),
just to name a few recent types.
Nanoflares are often attributed to magnetic energy
reconnection processes. Misaligned magnetic field lines
can break and reconnect, producing nanoflares in
avalanche-like SOC processes in form of fast and bursty
nanojets, escaping in opposite directions in an X-type
magnetic field configuration, and thus are interpreted
as reconnection-based coronal heating in action 
(Antolin et al.~2021). 
Energy distributions of nanoflares are found
to have a relatively steep slope ($\alpha_E=2.2 - 2.9$), and thus
appear to be in favor of dominant coronal heating by nanoflares 
(Ulyanov et al.~2019), except one abandons the unrealistic
flat-world 2-D geometry with a constant line-of-sight depth 
$h=const$) and replaces it with a more realistic 3-D geometry 
with an isotropic SOC model ($h=A^{1/2}$ and $V = Ah = A^{3/2}$).
A new type of small-scale EUV brightengings was introduced,
termed ``campfires'', which mostly occur at supergranular
boundaries and network junctions, but otherwise share 
similar statistics as nanoflares, $\alpha_E \approx 2-3$,
in the energy range of $E_{th} \approx 10^{23}-10^{26}$ erg
(Ulyanov et al.~2019; Alipour et al.~2022). 


\subsection{Energetics}

There are many ways to define energies in solar flares, which
tell us what energy dissipation mechanisms and what energy 
conversion mechanisms take place. 
The most comprehensive energetics of solar flares has been 
modeled in a series of 13 papers (labeled as Paper I, II, 
..., XIII), focusing on magnetic energies (Papers I, IX),
thermal energies (Paper II), nonthermal energies (Papers III,
VIII), CMEs (Papers IV, VI, VII), and various scaling laws
(Papers V, X, XI, XII), (Aschwanden 2020a, 2020b, 2020c, 2022c;
Aschwanden et al.~2021). 

A flow chart of solar flare energies is given in Fig.~(4), 
starting from the magnetic potential field energy $E_{\rm mag}$ 
and free energy $E_{\rm free}$, which feeds the 
primary energy dissipation processes 
(electron acceleration $E_{nt,e}$, ion acceleration $E_{nt,i}$,
direct heating $E_{dir}$, and launching of CME $E_{\rm CME}$), 
and secondary energy dissipation processes (thermal energy 
$E_{th}$, solar energetic particles $E_{\rm SEP}$, and bolometric
luminosity $E_{\rm bol}$), with radiative energies observed 
in white light $E_{\rm WL}$, soft X-rays, and EUV $E_{\rm rad}$,
(Aschwanden et al.~2017).

In the case of nonthermal energies,
for instance, electrons are accelerated in a collisionless 
plasma (such as in a magnetic reconnection region in the solar 
corona) and become thermalized 
when they enter the collisional plasma (by precipitation
into the solar transition region or chromosphere), a process 
that is described by the thick-target bremsstrahlung model). 
The energy associated with the dissipation of nonthermal 
electrons is called the {\sl nonthermal energy} $E_{nt}$.
It is useful to compare the nonthermal energy with
a reference energy, such as the free (magnetic) energy $E_{free}$,
which is the difference between the nonpotential and the
potential magnetic energy, 
\begin{equation}
	E_{free} = E_{np} - E_p =
	\int {B_{np}({\bf x})^2 - 
	      B_{p} ({\bf x})^2 
	\over 8\pi} dV ,
\end{equation}
Magnetic modeling yields the result that the mean dissipated 
flare energy in a solar flare amounts to $7\%\pm3\%$ of the
magnetic potential field energy, or $60\%\pm26\%$ of the free
energy $E_{free}$. Consequently, this result can be used to
predict flare magnitudes based on the potential
field of active regions before the flare occurs (Aschwanden 2019b).
The free (magnetic) energy constitutes
an upper limit for the energy dissipation in a flare.
The energy in nonthermal electrons, $E_{mag} \lapprox E_{free}$, 
is found to have a mean ratio of 
$E_{nt}/E_{mag}=51\%\pm17\%)$ (Aschwanden et al.~2017, Paper V),
$E_{nt}/E_{mag}=41\%\pm17\%)$ (Aschwanden 2016a, Paper III),
$E_{nt}/E_{mag}=57\%\pm08\%)$ (Aschwanden 2019a, Paper VIII).
$E_{nt}/E_{mag}=34\%\pm09\%)$ (Aschwanden 2019b, Paper IX).
The magnitudes of the flares selected here includes GOES
X-class and M-class events.
In summary, the nonthermal energy in electrons amounts to
about half of the free magnetic energy dissipated in a flare.
The nonthermal energy in electrons contains the largest 
contribution that is dissipated during a flare, while all 
other energy contributions are less important, such as the
energy in 
thermal ions          $(E_{ions}/E_{mag}   =17\%\pm17\%)$, 
thermal electrons     $(E_{th}/E_{mag}     = 8\%\pm13\%)$, 
direct (wave) heating $(E_{dir}/E_{mag}    = 7\%\pm17\%)$, or
CME kinetic energies  $(E_{\rm CME}/E_{mag}= 7\%\pm14\%)$ 
(Aschwanden 2017, Paper V). 


The free energy $E_{free}$ represents an upper limit of the
energy dissipated during a flare. The potential field energies 
in solar flares were found to vary in the range of
$E_p \approx 1\times 10^{31}-4 \times 10^{33}$ erg for 
GOES M- and X-class flares, while the free energy has a ratio 
of $E_{free} \approx 1\%-25\%$ (Aschwanden et al.~2014a).
For smaller flares, say GOES B- and C-class, which
we call nanoflares, the (multi-thermal) energy range reduces to 
$E_{diss}=10^{23}-10^{33}$ erg (Ulyanov et al.~2019),
which is linked to magnetic reconnection processes.  
The occurrence of helical twisting and braiding, which
is an expected magnetic topology, provides further direct
evidence for magnetic reconnection processes (Aschwanden 2019b).
Some techniques allow us to deduce the 3-D magnetic field even 
with ``blind'' stereoscopy (Aschwanden et al.~2015b).


\section{Solar Photosphere} 

\subsection{Size Distributions}

SOC phenomena detected in the solar photosphere are mostly
obtained from the 6173 \ang\ Fe I line. Some recent
statistics using HMI/SDO data were obtained by
Javaherian et al.~(2017), which exhibits broken
power law distributions with slopes of $(\alpha_1=2.24)$
and $(\alpha_2=4.41)$. The low slope value $\alpha_1$ is 
consistent with the SD-SOC prediction for 2-D areas,
$\alpha_A=(9/4)=2.25$ (Table 1), while the high slope value
$\alpha_2$ appears to be subject to finite system size
effects. The prediction of 2-D areas is indeed confirmed
by the applied detection algorithm, called {\sl Yet Another
Feature Tracking Algorithmn (YAFTA)}, which samples the
(projected) flare areas $A$, rather than (isotropic) 
volume-integrated 3-D flare fluxes. So, we have a 
predictive capability of the detection method-dependent 
(2-D vs. 3-D) size distribution.

The size distributions of photospheric SOC phenomena are
not always power law functions, as the broken power laws 
demonstrate (Javaherian et al.~2017). Alternative tested 
distributions include Weibull distributions of sunspot 
group areas (Shapoval et al.~2018), or Gaussian-plus-powelaw
(log-normal) size distributions (Aschwanden and Nhalil 2023a, 2023b). 


\subsection{Photospheric Magnetic Fields}

There are two fundamentally different methods to calculate
the magnetic field on the Sun, either by using Stokes polarimetry
in iron lines and force-free magnetic field extrapolation from
the photospheric boundary condition, or by forward-fitting of
parameterized loop projections using high-pass filtering of 
EUV images. A good agreement between the photospheric 
extrapolation and a coronal forward-fitting method has been 
accomplished for the magnetic field of Active Region 11158 
during the 2011 February 12-17 flares (Aschwanden et al.~2014b).
This test implies the equivalence of photospheric and coronal
magnetic fields, which is a fundamental assumption for 
the comparison of photospheric and coronal SOC phenomena.
In other words, regardless of the choice of magnetic data,
every 3-D magnetic field solution should converge to the
same unique magnetic field solution. In practice, the best
solutions were achieved for misalignment angles (betweeen
the projected magnetic and EUV coronal loops) of $\mu 
\approx 4^\circ$ (Aschwanden et al.~2014b, Table 3 therein).

The connection between the photospheric and the coronal 
magnetic field has been investigated with numerous methods,
such as estimating the rate of field line braiding in the
corona by photospheric flows (Candelaresi et al.~2018), or 
cellular automaton models that simulate avalanches of
magnetic flux ropes in the state of SOC (Wang et al.~2022).


\subsection{Multi-Fractality}

Volume estimates of gases or plasmas are expressed in the 
simplest way by Euclidean geometry, i.e., by means of a 
cube with the scaling of $V=L^3$ or $V=A h$. Such a notion 
may be adequate for a plasma environment with a high 
$\beta_{\rm plasma}$, 
\begin{equation}
	\beta_{\rm plasma} = {p_{\rm th} \over p_{\rm magn}} \ ,
\end{equation}
implying that the thermal pressure exceeds the magnetic
pressure.
For low $\beta_{\rm plasma} \ll 1$, however, the magnetic
pressure dominates and forces the plasma to move along the 
magnetic field. This has the consequence that the plasma
is distributed in near-parallel loops, with variable 
density and temperature inhomogeneities in perpendicular
direction to the local magnetic field, due to the lack of 
cross-field transport. This simple concept implies 
multi-fractal geometry in photospheric, chromospheric, 
and coronal plasma (Aschwanden and Peter 2017). 
The classical fractality is generally defined with a 
single fractal dimension (Eq.~6), (Mandelbrot 1977),
or more generally with 
a spectrum of fractal dimensions, called multi-fractals 
(see review by McAteer et al.~2016).
 
Giorgi et al.~(2015) investigated the question whether
measurements of fractal and multi-fractal parameters are
able to distinguish between flaring and nonflaring 
active regions, but found that C- and M-class flaring
active regions are indistinguishable, which is also true
for M-class and X-class active regions.
Alternatively, a multi-fractal segmentation technique
was used to detect the emergence of new magnetic flux,
which yielded the result of considerable variability of
areas with new magnetic flux and increase of complexity
on a characteristic time scale of $\approx 1-2$ h 
(Golovko and Sladkhutidinova 2015, 2018).


\subsection{Complex Networks}

The study of complex systems requires the analysis of network
theory (Gheibi et al.~2017). The hope is that changes in the
system may extract a pattern for (flare) prediction. For solar
flares, a graph theory has been employed to construct the complex
network. A network (graph) consists of nodes (vertices) and
edges (links). The graph can be simple, directed or
undirected, weighted or unweighted, regular, complete, scale-free,
or small-world. Some characteristics are degree distribution,
clustering coefficient, characteristic path length, and diameter.
Active regions act as hubs all over the network. 

A network
analysis of 14,395 active regions demonstrated that the degree
correlation has the characteristics of a disassortative network.
About 11\% of the large energetic M- and X-class flares 
occurred in the network hubs, which cover 3\% of the solar surface
(Gheibi et al.~2017). Another study applied a Hurst exponent 
analysis (0.8-0.9), a detrended fluctuation analysis, and 
a rescale range analysis to the time series. The behavior of the
clustering coefficient suggests that the active region network
is not a random network, but rather a SOC system with the
characteristic of a so-called small-world network 
(Daei et al.~2017).


\subsection{Photospheric Turbulence}

The general scenario emerged that the fully developed
turbulence in the convection zone generates
and transports magnetic flux tubes to the solar surface
(see review by Vlahos and Isliker 2016).
Photospheric turbulence has been invoked 
in super-granulation observations on spatial scales of $~30$ Mm 
(Rincon et al.~2017; Giannattasio et al.~2020),
in small-scale heating events simulated with the 
{\sl Bifrost} MHD code (Kanella and Gudiksen 2017),
in analytical models of convection-driven generation of
ubiquitous coronal waves on granular scales of $~1$ Mm
(Aschwanden 2018),
in preflare magnetic variability $~1-2$ days before flares
(Abramenko 2015, 2024), 
in thin photospheric flux tubes with 200-400 km diameter
(Abramenko and Yurchyshyn 2020),
in the L\'evy-walk (super-diffusion) characteristics 
of magnetic elements in the quiet Sun 
(Giannattasio et al.~2019; Giannattasio and Consolini 2021). 
The Bifrost simulations indicate power law size distributions
of the heating rate that are typical for SOC statistics
(Kanella and Gudiksen 2017).
While photospheric turbulence is not consistent with SOC behavior,
it may be more consistent with self-organization (without 
criticality) (Abramenko and Yurchyshyn 2020; Abramenko and Sulymanova 2024).
The largest disparity between SOC and turbulent systems is the
violation of time scale separation (in turbulent systems),
i.e., event durations are longer than the waiting times,
and thus subsequent events cannot be separated, and therefore
size distributions and power laws cannot be properly
measured.


\section{Solar Chromosphere}

The launch of the {\sl Interface Region Imaging Spectrograph (IRIS)}
in 2013 opened up our view on the chromosphere, sandwiched between
the photosphere beneath and the transition region and corona above.
Curvi-linear structures (fibrils, loop segments, spicules) are 
automatically detected with the {\sl Vertical Current Approximation
Nonlinear Force-Free Field (VCA-NLFFF)} code, which yields the reconstruction
of the chromospheric-coronal magnetic field with an accuracy (misalignment
angle) of $\mu \approx 4^\circ$ (Aschwanden et al.~2016c). The height
range of chromospheric features extends up to $h \approx 35,000$ km,
which is much larger than the expected hydrostatic scale height and this way
implies a dynamic atmosphere in ubiquitous non-equilibrium
(Aschwanden et al.~2016c).

Small-scale impulsive events observed with IRIS on the active Sun
reveal power-law energy distributions, with slopes of $\alpha_E
=1.80-2.07$ (Vilangot Nhalil et al.~2020), which is similar to
SOC parameters of coronal EUV brightenings or nanoflares.
The size distribution of energies is expected to be $\alpha_F=1.80$ 
for isotropic 3-D flux volumes, and $\alpha_A=2.33$ for 2-D areas
(Aschwanden 2022a, 2022b). Some new features were noted in chromospheric
SOC parameters, in particular a wider range of the fractal dimension: 
$D_A=1.21\pm$0.07 for photospheric granulation, 
$D_A=1.29\pm0.15$ for plages in the transition region, 
$D_A=1.54\pm0.16$ for sunspots in the transition region, 
$D_A=1.56\pm0.08$ for sunspots in EUV nanoflares, 
$D_A=1.76\pm0.14$ for large solar flares, and
$D_A=1.89\pm0.05$ for the largest X-class flares.
One expects low values $D_A \approx 1.0-1.5$ 
of the fractal dimension 
for sparse curvi-linear flow patterns, while high values 
of the fractal dimension are expected 
$D_A \approx 1.5-2.0$ for space-filling transport 
processes such as for the chromospheric evaporation model.
Thus, four groups of phenomena can be distinguished 
in the solar transition region and chromosphere, 
based on their size distributions of fractal areas 
and (radiative) energies: 
(i) Gaussian random noise in IRIS data,
(ii) spicular events in the plages and transition region,
(iii) salt-and-pepper small-scale magnetic structures, and
(iv) magnetic reconnection processes in flares and nanoflares
(Aschwanden and Nhalil 2023a, 2023b).


\section{Solar Corona}

The solar corona is permeated by so-called {\sl loops} or {\sl flux tubes}, 
which are nothing else than magnetic field lines that are filled with
heated plasma that is confined in low $\beta_{\rm plasma} \ll 1$ regions (where
the magnetic pressure dominates, and cross-field transport is inhibited).
These coronal flux tubes can be considered as SOC phenomena, because
they exhibit an exponential growth phase (when they become filled
and heated by chromospheric evaporation) and a fractal-diffusive decay 
phase (when they cool down by conductive loss and radiative loss).
A flare, in contrast, repesents a more complex and more dynamic 
multi-loop system that involves a change in magnetic topology.

The cross-section of coronal loops or flux tubes have a finite
diameter that varies, depending on the spatial resolution of the
observing telescope, over a large range from a maximum value of
$w_{max}\approx 10$ Mm down to a minimum value of $w_{min} \approx
100$ km (Aschwanden and Peter 2017). The latter value was a point
of contention, but its validity was proved by the Hi-C instrument,
which has the highest spatial resolution of 
$\Delta x=0.1^"=70$ km (Aschwanden and Peter 2017).

The magnetic field of a coronal loop can be approximated by
a semi-circular geometry in the case of a magnetic potential field,
or by a sigmoid geometry in the case of a nonpotential field.
The dynamic evolution of coronal loops has been studied by 
{\sl nonlinear force-free field (NLFFF)} codes 
(e.g., Wiegelmann and Sakurai 2012, Aschwanden 2016b).
The nonlinear force-free fields typically exhibit
helical twists, which is manifested in strong vertical currents,
which can be characterized with the helical twisting number
(which was found to have a mean of 
$N_{\rm twist}=\Phi/360^0=0.14\pm0.03$), or
with the braiding linkage number 
(Aschwanden 2019d; Pontin and Hornig 2015).
The evolution of a braided coronal loop, using a model with
multiple threads, twisting and braiding around each other,
has been simulated and showed that the reconnected
field lines evolve into a SOC state (Berger et al.~2015).
Often, only one single thread needs to be unstable to
start an avalanche, even when the others are below
marginal stability (Hood et al.~2016).

The smallest coronal strutures are called 
{\sl extreme ultraviolet coronal bright points}
(Guerreiro et al.~2015;
Chesny et al.~2016; Alipour et al.~2022),
which occur in the Quiet-Sun (away from active
regions), with a negative correlation to the
number of sunspots (Alipour and Safari 2015), 
and they tend to display a single-loop geometry, 
or sigmoid geometry (Chesny et al.~2016),
similar to nanoflares.
Numerical simulations suggest that braiding-associated
reconnection in the corona can be understood in terms
of a SOC model driven by convective rotational motions
at their footpoints, reproducing power law size distributions
of $\alpha_E=1.76\pm0.02$ (Knizhnik et al.~2018).


\section{Heliosphere}

\subsection{Coronal Mass Ejections (CME)}

The probably most comprehensive statistical analysis of
{\sl coronal mass ejections (CME)} has been carried out
from LASCO/SOHO coronagraph data, covering two complete
solar cycles 23 and 24, using the 5 catalogs CDAW, 
ARTEMIS CACTus, SEEDS, and CORIMP (Lamy et al.~2019).
The temporal evolutions and distributions of their
properties have been characterized for occurrence
and mass rate, waiting time, periodicity, angular
width, latitude, speed, acceleration and kinetic
energy. The following solar activity indices are
found to be good proxies for the magnitude of CMEs:
the radio flux at 10.7 cm, the international sunspot
number, the sunspot area, and the total magnetic field.
The distribution of kinetic energy follows a log-normal
law, and that of angular widths follows an exponential
law, implying that they are random and independent.
The distribution of waiting time was found to have
a long power law tail with a power law slope of
$\alpha_{\rm LASCO} \approx 2.0-3.0$ (Lamy et al.~2019), 
which is consistent the theoretical prediction of the
SOC waiting time, $\alpha_{\rm WTD} \approx 2.0-3.0$ 
(Aschwanden and McTiernan 2010). New trends
concentrate on perfecting the detection algorithms of CMEs,
which strongly depends on the viewing geometry, from 
behind the limb (halo-CMEs) to the center of the disk, 
and should lead to reconciliation of the various CME catalogs. 
Due to the occultation of the disk in the case of 
coronagraph white-light observations, 
supporting disk observations (e.g., in EUV, soft X-rays,
H-$\alpha$ and radio) are needed in order to 
reconstruct the full 3-D geometry and kinematics of CMEs
(Georgoulis et al.~2019). Moreover, different physical 
SOC parameters play a role in the statistics of CME energies, 
such as the kinetic energy, the gravitational potential
energy (Aschwanden 2016a, 2017), and the aerodynamic drag force,
which adds an initial boost to a launched CME due to the
ambient solar wind speed (Aschwanden and Gopalswamy 2019).


\subsection{Solar Wind}

The {\sl Parker Solar Probe} (Fox et al.~2016), 
launched on 2018 August 12, 
circling the Sun in gradually shrinking elliptical orbits,
ultimately approaching the Sun at a closest distance of 
$d \approx 10 R_{\odot}$, opened a powerful new opportunity 
to study the solar wind in-situ at distances of $d=9.8-35.7 
R_{\odot}$. Earlier measurements with ACE and Ulysses
probed the solar wind much further away from the Sun,
at distances of 0.3 AU to $\ge 8.0$ AU (e.g., Chen and Hu 2022).

An intriguing discovery is the so-called phenomenon of
``switchbacks'', which show up as omnipresent swift reversals
of the magnetic field, which otherwise follows the mostly
radial Parker spiral, but exhibits deflection angles over
the entire angular range of $0^\circ-180^\circ$ with respect to
the Parker spiral (Dudok et Wit et al.~2020). 
There is a continuum of deflections
that could typically result from a fractal distribution of
kinked magnetic flux tubes. Waiting times and residence times
suggest two regimes, a switchback component, and a quiescent
solar wind component, in agreement with classical (pristine)
solar wind background turbulence.
The fact that the sunspot number is closely correlated with
the waiting time power law slope $\alpha_{\rm WTD}$
indicates a dependence of the detection threshold of events 
on the instrumental sensitivity, similar as found
for other SOC datasets (Aschwanden and Dudok de Wit 2021).

Another study (Chen and Hu 2022) with data from the 
Parker Solar Probe
presents measurements of some $\approx 3 \times 10^3$
small-scale magnetic flux ropes, detected with the
Grad-Shafranov algorithm, which have durations 
from 10 s to $\lapprox 1$ hr and cross-field sizes
from a few 100 km to $10^{-3}$ AU. Their size distributions
show power law functions with a slope of $\alpha_T=1.76$ for
durations, and $\alpha_V=1.87$ for (volume) scale sizes,
which are roughly consistent with the size distributions
predicted by the FD-SOC model (Table 1).

A new trend in modeling the solar wind concerns the 
waiting time distribution, for which the power law 
distribution slope is (Aschwanden and Johnson 2021),
\begin{equation}
	\alpha_{\rm WTD} = 2 + {1 \over p} \ ,
\end{equation}
while $p$ is the polynomial exponent for the time-dependent 
flare rate $\lambda(t)$, 
\begin{equation}
	\lambda(t) = \lambda_0 \left({t \over T/2}\right)^p \ .
\end{equation}
The analytical derivation of the exact solution of the
waiting distribution is given in Aschwanden (2025a) and
references therein.
In the linear limit of $p=1$ we expect a waiting time power
law slope of $\alpha_{\rm WTD}=2+(1/p)=3$. For a moderate
nonlinear value of $p=2$, we expect  
$\alpha_{\rm WTD}=2+(1/p)=2.5$, and for an extremely
nonlinear case of $p \mapsto \infty$ we expect 
$\alpha_{\rm WTD}=2+(1/p)=2$.
Thus for the range from $p=1$ to $p \mapsto \infty$ we expect 
a power law range of $\alpha_{\rm WTD}=2-3$, depending on the 
time-dependent (non-stationary) flare rate, which appears to 
be in agreement with solar flares, CMEs, geomagnetic storms, 
and substorms (Nurhan et al.~2021). 
The same range $\alpha_{\rm WTD}=2-3$ was been established from
the waiting time distributions earlier (Aschwanden and McTiernan 2010).
This theoretical result may be compared with the observations
of the switchbacks statistics in future work, in order to
answer the question whether solar wind switchbacks or magnetic 
flux ropes qualify as SOC phenomenona.


\subsection{Solar-Energetic Particle Events (SEP)}

There is a substantial controversy whether 
{\sl Solar-Energetic Particle (SEP)} events
can be interpreted as a SOC phenomenon or not.
The disparity of power law slopes between
SEPs and solar flares has already been
pointed out by Hudson (1978), who suggested
three possibilities to explain the difference 
between the observed value ($\alpha_P=1.15$)
versus the theoretically expected value ($\alpha_F=1.8$).
Although the sizes of SEP events display more
or less a power law distribution function,
the correponding power law slopes have been 
reported to have significantly lower power 
slopes of $\alpha_P \approx 1.2-1.4$
(Aschwanden 2011, 2025a, and references therein),
rather than those measured in solar flare
hard X-rays or soft X-rays, with
slopes of $\alpha_P \approx 1.8-2.0$.
Hudson (1978) suggested 3 possibilities 
to explain the
difference between $\alpha_{\rm SEP}=1.15$
and $\alpha_{\rm HXR}=1.80$: (i) flare with
SEP events differ fundamentally from ordinary
flares, (ii) flares with SEP events are merely
the large end of the flare distribution, and
(iii) flares with SEP events have a threshold
behavior.

Cliver et al.~(2012) and Cliver and D'Huys (2018) 
suggested that the flatter
size distribution of solar energetic proton 
events relative to that of flare soft X-ray
events is primarily due to the fact that SEP
flares are an energetic subset of all flares.
In contrast, Kahler (2013) argues that the 
spatial dimensionality (with Euclidean dimension 
$d=1$) of the generalized FD-SOC model explains 
the flatter size distribution of SEP events
($\alpha_V=1.0$), while the standard
FD-SOC model (with Euclidean dimension $d=3$)
explains the steeper slope of solar flare events
($\alpha_V=1.8$). The implication of this argument
is then what two different physical mechanisms
require to switch on different dimensionalities
$d=2$ and $d=3$? Kahler (2013) suggests that
the dimensionality $d=2$ is produced by SEPs
that are accelerated at 2-D shock surfaces, 
while the dimensionality $d=3$ is produced 
during solar flares in soft X-ray-emitting 
3-D (isotropic) volumes of heated plasma.  


\section{Magnetosphere}

In recent literature, self-organized 
criticality (SOC) has been invoked for 
a number of magnetospheric processes, 
such as the geomagnetic AE index of the
interplanetary field (Souza et al.~2016),
magnetospheric substorms (Bolzan 2018),
the solar wind-magnetoshperic coupling
(Banerjee et al.~2019), Carrington-like
geomagnetic storms (Morina et al.~2019), or
magnetospheric superstorms (Love 2020). 

Souza et al.~(2016) focused
on the interplanetary and geomagnetic
characteristics of {\sl High-Intensity
Long-Duration Continiuous AE activity 
(HLDCAA)} events and interpreted the
periodicities in terms of SOC phenomena,
based on the capacity of releasing large
amounts of energy in short time intervals,
which, however, is at odds with standard SOC 
behavior, for which no periodicities are 
expected.

Bolzan (2018) suggested that the responses
of the Earth magnetic field during substorms
exhibit a number of characteristic SOC features,
such as power law spectra of fluctuations on
different scales and signatures of low 
effective dimensions.

A SOC approach to study the solar wind-
magnetosphere energy coupling was presented
by Banerjee et al.~(2019), using a cellular
automaton model that produces the real-time
Dst index and AE index series. Agreement
between the numerically simulated AE index
and the observed AE index supports the SOC
interpretation.

A practical application of SOC theory is the
prediction of the largest geomagnetic event 
based on the modeled SOC size distribution.
Morina et al.~(2019) estimated the probability
of a Carrington-like geomagnetic storm and
came down with a range of $0.46\%-1.88\%$
(with 95\% confidence level), which is a much lower
value than quoted in literature. One major
uncertainty is the finite system size effect,
which truncates the power law size distribution
at the largest events.

In order to pin down the detailed size distribution
of a SOC dataset of magnetic superstorm intensities,
various methods have been used, such as log-normal,
upper limit log-normal, generalized Pareto, and
generalized extreme-value models have been employed.
All methods provide good representations of the
data, but no clear indication is given which model 
is best (Love 2020). This outcome is consistent
with the fact that log-normal size distribution 
converges towards a power law at the upper end.


\section{Planetary Systems}

\subsection{Asteroids and Small Bodies}

When we look at the size distribution 
of various small bodies
in our solar system (Aschwanden et al.~2016b),
we find similar size distributions for
asteroids, near-Earth objects, Jovian Trojans, 
or Neptune Trojans (Fig.~5),
with a typical cumulative power law slope of 
$\alpha_L^{cum} \approx 2.0$, which translates into
a differential size distribution with slope
$\alpha_L \approx \alpha_L^{cum}+1=2+1=3$. 
Hence, the size distribution $N(L)$ of length
scales is consistent with the scale-free assumption
of SOC statistics,
\begin{equation}
	N(L) dL \propto L^{-3} \ dL .
\end{equation}
While the size distribution of small bodies in our 
solar system is the result of numerous collisional
processes, other solar system phenomena with collisional
or fragmentation processes are expected
to follow the same SOC statistics, such as 
Saturn ring particles, lunar craters, or the Kuijper belt
(Fig.~5), which is indeed the case, given the observed
adherence to the scaling $\alpha_L^{cum} \approx 2.0$. 
Lunar craters are created by the collision of an asteroid
with the lunar surface, leaving a circular impact crater
behind and a radially propagating shock wave. 
Similarly, asteroid impacts occur on the Earth as well
and leave craters behind, such as the event about 66 million
years ago that created the Chicxulub crater in Mexico
and caused the extinction of the dinosaours and 75\%
of all species on Earth.
The process of crater impacts is not too much different 
from the collision of tectonic plates during earthquakes,
which also produces a spherical shock wave 
originating at the epicenter,
which is considered to be a well-established SOC process.

The cratering in our solar system is found on Earth, the Moon,
Mars, or Mercury. The variabillity of the impact rate
varied drastically during the lifetime of our solar system.
Both the Moon and the Earth were subject to intense
bombardment between 4.0 and 3.7 billion years ago,
which was the final stage of the sweep-up of debris
left over from the formation of the solar system
(Bottke et al.~2012). The impact rate at that time
was thousands of times higher than it is today.

\subsection{Saturnian Ring Systems}

The size distribution of spatial scales has been
measured with {\sl Voyager I} by means of radio
occultation observations, yielding distributions
of $N(L) \approx L^{-3}$. It appears that the size
distributions of asteroids, lunar craters, 
meteorites, and Saturnian ring particles can all
be explained by the same (FD-SOC) model.

Saturnian ring systems are also found around other large
planets with sufficiently strong gravitational field to 
keep numerous moons, rings, and ringlets in trapped
orbits. Theories
about the origin of Saturn's ring range from
nebular material left over from the formation
of Saturn itself, collisional fragmentation,
to the tidal disruption of a former moon.
The critical threshold in these systems is
apparently given by the balance between the
local self-gravity and external gravity
disturbances. Although resonant orbits are
likely to play a role, based on the harmonic
ratios observed in planet (or exoplanet) orbits,
their collisional avalanches clump in zones of
nonresonant orbits. The appeal of the FD-SOC
model is the simple way to predict the final
size distribution of ringlets that result in the end,
which cannot be predicted by celestial mechanics
of chaos theory.  

\subsection{Terrestrial Gamma-Ray Flashes}

{\sl Terrestrial gamma-ray flashes (TGF)} are
gamma-ray bursts of terrestrial origin that have
been discovered with the {\sl Burst and Transient
Experiment (BATSE)} onboard the {\sl Compton
Gamma Ray Observatory (CGRO)} and have been
studied with RHESSI, Fermi, and AGILE since.
These TGF bursts are produced by high-energy
photons of energy $>100$ keV and last up to
a few milliseconds. They have been associated
with strong thunderstorms mostly concentrated
in the Earth's equatorial and tropical regions
at a typical height of 15-20 km 
(Fishman et al.~1994). 
The physical interpretation is that the TGF
bursts are produced by bremsstrahlung of
high-energetic electrons that were accelerated
in large electric potential drops within
thunderstorms. However, the gamma-rays
produced in thunderstorms (at 5 km) cannot
readily propagate to higher altitudes due to
atmospheric absorption. A mechanism for the
generation of gamma rays that can reach the
satellite-borne instruments is through
excitation of whistler waves by the relativsitic
electrons generated in the thunderstorms.
The whistler waves form a channel by nonlinear
self-focusing, and the relativistic electrons
propagate in this channel to higher altitudes
(30 km). The gamma-rays generated at this altitude
can escape the atmosphere and thus account for
the BATSE/CGRO results (Fishman et al.~1994).

The size distribution of TGF fluence was found
to produce a power law with slopes of
$\alpha_E = 1.3-1.7$, after a roll-over 
of the RHESSI lower detection threshold is applied 
(Ostgaard et al.~2012).
This observational value matches the theoretically
predicted value of $\alpha_E=1.67$ of the fluence
in 3-D space (Eq.~12).

\section{Stellar Flares}

\subsection{Kepler Data}

The Kepler space telescope (Borucki et al.~2010),	
launched on 2009 March 7, was designed to survey the
stars and planets in the Milky Way galaxy, but
is equally well suited to study stellar flare events,
which stick out of the background of rotation-modulated 
light curves in optical (white-light) wavelengths.
Extensive datasets and catalogs of stellar 
flares were generated from recent Kepler data 
(Wu et al.~2015; Lurie et al.~2015;
Davenport 2016; Silverberg et al.~2016; 
Svanda and Karlicky 2016; Van Doorsselaere et al.~2017;
Gizis et al.~2017a, 2017b; Aschwanden 2019e; 
Aschwanden and G\"udel 2021; Kuznetsov and Kolotkov 2021).      
The largest numbers of detected stellar flare events amount to 
n=851,168 events in 4,041 candiatate flare stars (Davenport 2016),
n=188,837 events in the Kepler field-of-view during the quarter Q15, 
n=162,262 events on 3,422 flare stars (Yang and Liu 2019), and 
n= 16,850 events on 6,662 flare stars (Van Doorsselaere et al.~2017). 
A synthesized size distribution of solar nanoflares, solar large
flares, and stellar flares is depicted in Fig.~(6), which demonstrates
the universality of self-consistent power law slopes over 13 orders 
of magnitude.

The detection rate of stellar flares obtained with
other instruments than Keper or TESS, 
such as EXOSAT, EUVE, HSP/HST, XMM-Newton,
are at least two orders of magnitude 
smaller and generally do not exceed $n \approx 10^2$ events.
However, one can not use the large-number statistics
alone to extract the most accurate values of the size
distributions, since systematic errors in the detection
methods can easily cause larger systematic errors than
the statistical errors. Moreover, the size distributions
may also depend on the stellar spectral type, in which 
case we have to sample events with a given stellar type, 
a given observing instrument, or an other selected criterion 
separately. 

Regarding different observational instruments (in the
wavebands UBV, XUV, Kepler, TESS), one finds power law 
slopes of (Aschwanden and G\"udel 2021)
\begin{equation}
        \alpha_{\rm F,UBV} =1.80 \pm 0.19 \ ,
\end{equation}
\begin{equation}
        \alpha_{\rm F,XUV} =2.08 \pm 0.37 \ ,
\end{equation}
\begin{equation}
        \alpha_{\rm F,Kepler} =1.93 \pm 0.40 \ ,
\end{equation}  
\begin{equation}
        \alpha_{\rm F,TESS} =2.16 \pm 0.10 \ ,
\end{equation}  
which is most consistent with the theoretical prediction
($\alpha_F=1.80$, Table 1) of the FC-SOC model for the 
UBV waveband.

Regarding the dependence of the power law slope on the
stellar spectral type, we find 
(Aschwanden and G\"udel 2021),
\begin{equation}
        \alpha_{\rm F,G-type} =2.10 \pm 0.21 \ ,
\end{equation}
\begin{equation}
        \alpha_{\rm F,F-type} =2.36 \pm 0.25 \ ,
\end{equation}
\begin{equation}
        \alpha_{\rm F,M-type} =1.89 \pm 0.35 \ ,
\end{equation}
\begin{equation}
        \alpha_{\rm F,K-type} =2.02 \pm 0.33 \ ,
\end{equation}
\begin{equation}
        \alpha_{\rm F,Giants} =1.90 \pm 0.10 \ ,
\end{equation}
\begin{equation}
        \alpha_{\rm F,A-type} =1.12 \pm 0.08 \ ,
\end{equation}
which contains 6 flare star types across the Hertzsprung-Russell
(H-R) diagram 
(Balona 2015).

Here we have an obvious outlier, namely the A-type star,
which is not understood at this time. 
For A-type stars, the total energy flux density was
found at least 4 times smaller than for G stars,
which may explain the lack of hot coronae on A-type stars
(Svanda and Karlicky 2016). 
The fact that some A stars were found to exhibit flaring 
activity is surprising, because these stars, according to 
stellar evolution theory, are not believed to have an 
outer convection zone. Thus, it is inconclusive whether 
A stars can flare or not (Balona 2015;
Van Doorsselaere et al.~2017). 

All other stellar spectral types
display mean values close to the prediction of the
FD-SOC model, $\alpha_F=1.80$ 
(Aschwanden and G\"udel 2021).
There is an overall trend that most stellar size distributions
are slightly steeper than predicted, which is likely to be a
finite system size effect, limited by the maxiumum starspot size. 
Another study, however, implies
that Dragon-King (extreme)  events are relatively rare
in both solar and stellar flares, i.e., 4 out of 25 cases
(Aschwanden 2019e).

\subsection{TESS Data}

The {\sl Transiting Exoplanet Survey Satellite (TESS)},
launched on 2018 April 18, provides data on stellar
flare detections also, amounting to 
1216 superflares on 400 solar-type stars (Tu et al.~2020); 
140,000 stellar flares on n=25,000 stars (Pietras et al.~2022); and 
$n \approx 10^6$ flaring events on $n \approx 10^5$ stars observed
by TESS at a 2-minute cadence (Feinstein et al.~2022).
TESS data revealed a higher superflare frequency 
distribution than Kepler, probably because the
majority of the TESS-detected solar-type stars are 
rapidly-rotating stars, but otherwise the 
power law slopes are similar to Kepler-detected
events, with $\alpha_{\rm F} = 2.16 \pm 0.10$ 
(Tu et al.~(2020). Stars with shorter rotation periods
tend to have larger power law slope $\alpha_E$ values
(Wu et al.~2015), which could explain the observed
slightly steeper power laws than predicted by the
FD-SOC model.

\subsection{Superflares}

The extensive solar and stellar datasets allow us to 
synthesize the solar and the stellar size distributions
into a common synthesized distribution
(Maehara et al.~2015, 2017;
Notsu et al.~2015a, 2015b;
Aschwanden et al.~2016b; 
Aschwanden and Schrijver 2025), 
which yields flare energies within the ranges listed in Table 2.
The physical unit of the flare energy was estimated by 
a mean magnetic energy of
\begin{equation}
	E_{magn}= \left( B^2 L^3 \over 8\pi \right) ,  
\end{equation}
with $B \approx 10^3$ G (Maehara et al.~2015), 
since it cannot be measured directly from stellar data.

Flare energy estimates without magnetic field strength
information have also been quantified with the 
Stefan-Boltzmann law (with constant $\sigma_{sb}$),
\begin{equation}
	E_{flare} = \int (4\pi R_*^2) \sigma_{sb} T_{\rm eff}^4
	F_{flare}(t) \ dt \ ,
\end{equation}
where $F_{flare}(t)$ is the normalized stellar flare flux,
$R_*$ is the stellar radius, and $T_{\rm eff}$ is the 
effective temperature (Wu et al.~2015).

Of particular interest is the upper limit of stellar
flares, a category that is also called superflares.
The uppermost values of stellar flare energies has
been quoted as 
$E_{max}=10^{36}$    erg (Maehara et al.~2015; 2017),
$E_{max}=10^{36.5}$  erg (Davenport et al.~2016), 
$E_{max}=10^{36.87}$ erg (Balona 2015), 
$E_{max}=10^{37}$    erg (Aschwanden and  Schrijver 2025), and
$E_{max}=10^{37.3}$  erg (Wu et al.~2015).

The statistics of superflares has also been tested with
both the nonstationary Poisson function and the Weibull
function, and it was found that relatively weak flares
are persistently correlated, while the production of
superflares is closer to a stochastic process, 
statistically (Li et al.~2018a).

\subsection{Stellar Rotation}

The surface magnetic field strength decreases over the life time
of a star, due to the steady loss of angular momentum, which
slows down the internal dynamo. Older, slowly-rotating stars
like our Sun exhibit smaller and fewer sunspots, while 
younger rapidly-rotating stars can produce long-lived
and large starspots and stellar flares. Saturation of flare activity
is estimated below a Rossby number of $R \approx 0.03$,
based on a Kepler catalog of stellar flares (Davenport 2016;
Balona 2019).

Avalanches and the distribution of reconnection events in
magnetized circumstellar disks have been modeled by
Fatuzzo et al. (2023)

\section{Galactic and Extra-Galactic Systems}

\subsection{Pulsar Glitches}

A pulsar is a highly magnetized, rapidly-rotating neutron star
that emits a beam of electro-magnetic radiation. The beamed
emission is aligned with the magnetic axis, so that we observe
rotationally-modulated pulses whenever the beam axis points
to the Earth line-of-sight during each period of its rapid
rotation, similar to the rotating beam of a light house beacon.
Besides the regular periodic pulses on time scales of
milliseconds, measured with high accuracy, there occur
sporadic unpinning glitches in pulse amplitudes and
frequency shifts, probably caused by sporadic unpinning
of vortices that transfer momentum to the crust
(Warszawski and Melatos 2019).

There are a lot of enigmatic observations of pulsar 
glitches.  The glitches of pulsars have been found 
to show some power law behavior, and thus have been 
associated with SOC systems, although the power law 
slopes themselves scatter over large ranges (see 
Table 14 in Aschwanden et al.~2016b), and thus are 
likely to be affected by small-number statistics 
and finite system size effects. Tests of the 
universality of SOC phenomena in galactic, 
extra-galactic, and black hole systems are mostly 
consistent with SOC theory, while pulsar glitches 
are not (Aschwanden and Gogus 2025; Aschwanden 2025a).

Kepler K2 observations of the transitional millisecond
pulsar PSR J1023+0038, however, yield much improved 
power law slopes of fluences, $\alpha_E=1.7$ 
(Kennedy et al.~2018), which agrees
well with the theoretical FD-SOC model, $\alpha_E=1.67$.
The flaring behavior of the pulsar PSR J1023+0038 was
intepreted in terms of a SOC mechanism,
most likely related to the build-up and 
release of mass at the inner edge of the accretion disk.
The neutron star rotates with a millisecond period
and is thought to be the descendant of a low-mass
binary, having been spun-up by accreting matter from
a companion star (Kennedy et al.~2018).

Pulsar glitches revealed suprising observational details.
Dual emission modes were detected where synchronous
X-ray/radio emission appeared to switch between a 
radio-bright mode and a radio-quiet mode
(Hermsen et al.~2018).
A comparative analysis of scale-invariant phenomena in 
glitching pulsars reveals that some pulsars exhibit 
spin-ups as well as spin-downs, also called anti-glitches, 
(Gao and Wei 2024). One wonders whether there is a
two-way mass transfer in an accreting binary star system.

New trends of theoretical modeling of pulsar glitches
include a combination of analytical modeling and
numerical simulations, including neutron stars with
persistent emission of current quadrupole gravitational 
radiation (Melatos et al.~2015), numerical Bayesian 
inference of glitch identification (Yu and Liu 2017),
rotating self-gravitating Bose-Einstein condensates 
with a crust (Verma et al.~2022), a neutron-superfluid 
vortices and proton-superconductor flux tube 
model for pulsar glitches (Shukla et al.~2024), and 
quantum-mechanical simulations of neutron star 
rotational glitches (Melatos et al.~2020).

\subsection{Gamma-Ray Bursts (GRB)}

The first electro-magnetic signal of a {\sl gamma-ray burst (GRB)}
is the so-called $\gamma$-ray prompt emission, followed by the 
early X-ray afterglow on a timescale from minutes to hours.
The most common interpretation of GRBs is associated with
the core collapse of a massive star, rather than external 
shocks, but the radiation process is not understood.

X-ray flares are generally supposed to be produced
by later activities of the central engine, and may
share a similar physical origin with the prompt
emission of GRBs. Analyzing data from Swift and
BATSE/CGRO reveal size distributions 
of peak fluxes, durations, fluences, peak luminosities
similar to solar hard X-ray flares 
(Wang et al.~2015; Yi et al.~2016, Lyu et al.~2020, 
2021; Wei 2023; Li and Yang 2023; Li et al.~2023;
Maccary et al. 2024), although some of the power law 
fits reveal deviations from predicted power laws, likely 
due to threshold effects and finite system size effects.

In order to further test the SOC interpretation, the statistics
of {\sl waiting time distribution (WTD)} has been employed.
It was found that prompt $\gamma$-ray and early X-ray 
afterglow emission are characterized by a bursty stochastic 
process, interspersed with long quiescent time intervals 
(Guidorzi et al.~2015).  

The spatial distribution of GRBs is isotropic 
(in the plane-of-sky), and thus is believed to have 
a cosmological origin, being located
at extra-galactic distances (Piran 1992; Harko et al.~2015).  
On the theoretical side, the SOC behavior of GRBs has 
been numerically simulated with a 1-D cellular automaton 
model, using a simplified MHD induction equation
(Harko et al.~2015).

The brightest GRB event observed so far is the very 
high-energy (TeV photons) afterglow in GRB 221009A,
which exhibits some SOC behavior (nonstationary WTD,
power laws, Tsallis q-Gaussian distributions), but
show also significant deviations from the predicted
power laws (Zhang et al.~2025). Nevertheless,
scale invariance is evidenced by fulfilling the
q-Gaussian distribution statistics (Wei 2023).

\subsection{Soft Gamma Ray Repeaters (SGR)}

{\sl Gamma Ray Bursts (GRBs)} were detected with CGRO, RXTE,
and ICE in the hard X-ray range of $\approx 20-40$ keV
(a wavelength that is also called soft-gamma rays).
GRB events that show repeated detections from the same
spatial source location have been dubbed {\sl Soft
Gamma Ray Repeaters (SGRs)}. 

The probably most analyzed
SGR event is observed in the gamma-ray source J1935+2154 
(Xiao et al.~2023a, 2023b; 2024a, 2024b, 2024c, 2024d, 2024e; 
Zhang et al.~2023; Xie et al.~2024; Li et al.~2024).
SGR J1935+2154 is one of the most active magnetars 
since its discovery in 2014.

The size distributions of fluctuations in duration,
waiting time, peak count rate, and total counts of
pulses within bursts from the two active and prolific
magnetars SGR J1935+2154 and SGR J1550-5418 have
been sampled from Fermi/GBM, Insight-HXMT, and GECAM
and were found to be similar to other SOC statistics 
(Zhang et al.~2023),
which was interpreted in terms of a common radiation
process for the two magnetars (Xiao et al.~2024d). 
The differential size distributions
is found to have a power law slopes of 
$\alpha_F=2.26\pm0.11$ for SGR J1550+5418, and 
$\alpha_F=2.05\pm0.07$ for SGR J1935+2154, 
which is close to the theoretical expectation
$\alpha_F=1.80$ of the FD-SOC model.

\subsection{Supergiant Fast X-Ray Transients}

Supergiant fast X-ray transients are a sub-class of 
high mass X-ray binaries, in which the compact object 
is a neutron star and is accreting materials from 
its non-degenerate companion star (Zhang et al.~2022).
The fact that the size distributions exhibit power law 
slopes in the range of $\alpha_F \approx 1.2-1.9$ for 
luminosities, is interpreted in terms of a SOC process
(Zhang et al.~2022). However, the dynamic (fitting) range 
in this study is often less than one decade, which makes
the power law fitting prone to small-number statistics 
and finite system size effects.

\subsection{Fast Transient Radio Bursts (FRB)}

{\sl Fast Radio Burst (FRB)s} are astrophysical radio 
flashes of unknown origin, with durations of 
milliseconds and large dispersion measures, i.e.,
(integrated column density of free electrons
between source and telescope).  
Their (radio frequency) dispersal arrival times
suggest an extra-galactic origin and imply radio
luminosities orders-of-magnitude larger than any
other kind of known short-duration transient. 
The only known repeating {\sl Fast Radio Burst (FRB)} 
source is FRB 121102, which has been observed with 
the Green Bank Telescope and the Arecibo Observatory 
(Spitler et al.~2016; Scholz et al.~2016;
Chatterjee et al.~2017). 
 
Studying a sample of FRBs observed with the radio 
telescope ASKAP yielded the size distribution of
the fluence $dN(E)/dE \propto E^{-\alpha_E}$
with $\alpha_E=1.7$ and dispersion measure (DM)
$N(>DM) \propto DM^{5-2\alpha_E}$. Such power law
slopes are natural consequences of SOC
(Lu and Piro 2019).   

A sample of 41 bursts from the source FRB 121102 was
detected with the Arecibo telescope at 1.4 GHz,
which yield a size distribution of $\alpha_F=1.8\pm0.3$
and perfectly matches the theoretical prediction of the
FD-SOC model, $\alpha_F=1.80$ (Gourdji et al.~2019).

Energy and waiting time distributions in a large
sample of 1652 bursts yields a power law slope of
$\alpha_F=1.86\pm0.02$, which is close to the theoretical
prediction of the FD-SOC model $\alpha_F=1.80$.
The power law slope is $\alpha_F=1.70\pm0.03$ for
early bursts, but much steeper for later bursts,
$\alpha_F=2.60\pm0.15$
(Zhang et al.~2021). 

The recent discovery of a galactic FRB associated 
with a hard X-ray burst from the soft gamma repeater 
(SGR J1935+2154) has established the magnetar origin of at 
least some FRBs. The scale invariance in both the
high-energy component of FRB 121102 and SGRs can be well
explained within the same physical framework of
fractal-diffusive SOC systems
(Wei et al.~2021).

Observed cumulative distributions can be interpreted in terms
of SOC theory, by a possible magnetic reconnection 
scenario in magnetars. The obtained power law slopes are
$\alpha_F=1.57-1.63$ for photon counts,
$\alpha_D=1.99-2.07$ for durationts,
$\alpha_{\rm WTD}=1.95-1.97$ for waiting times,
(Wang et al.~2023a, 2023b).

Power-law distribution and scale-invariant structure
are determined from the first CHIME/FRB fast radio burst
Catalog. A selected sample of 190 on-off FRBs
was found to agree with predictions of SOC systems
(Wang et al.~2023a, 3023b).

Similar scale-invariant properties can be found 
in PSR B1737–30 glitches, implying an underlying 
association between the origins of repeating FRBs 
and pulsar glitches. These statistical features 
can be well understood within the same physical
framework of self-organized criticality system
(Gao and Wei 2024). 

Statistical properties of energy and waiting time
carry essential information about the source of
repeating fast radio bursts.  All these statistical 
properties imply that, although the time series of 
repeating FRBs seems to be irregular, that they are 
not completely random, similar to the features of 
self-organized criticality
(Sang and Lin (2024).

\subsection{Magnetars}

Magnetars are observed as peculiar, burst-active X-ray
pulsars, the anomalous X-ray pulsars (AXPs), and the
soft-gamma repeaters (SGRs) (Turolla et al.~2015).
Kouveliotou et al.~(1999) discovered a soft gamma-ray
repeater (SGR 1900+14) associated with a magnetar.
The SGR 1900+14 exhibits a spin-down rate of
$1.1\times 10^{-10}$ s$^{-1}$ and is attributed to
magnetic dipole radiation, possibly accelerated by
a quiescent flux, yielding a pulsar magnetic field
of $B=(2-8)\times 10^{14}$ [G], and confirms that
SGRs are magnetars (Kouveliotou et al.~1999).

Recently, FRB 200428 was detected in temporal 
coincidence with a hard X-ray flare from the 
Galactic magnetar SGR 1935+2154, which supports 
that at least some FRBs are from magnetar activity.
(Wang et al.~2021)

A Bayesian model was developed for magnetars. It was
found that the sub-burst variability does not conform
to SOC models, and thus an amplified cascade model for 
the different trigger mechanisms has been suggested
(Huppenkothen et al.~2015).    

More recently, further magnetar observations were reported 
based on the statistical properties of magnetar bursts, 
and FRB 121102 shows approximate agreement with SOC 
parameters of power law slope in energy, duratation, 
and waiting times, but the size distributions display 
significant deviations from ideal power law functions 
(Enoto et al. 2017;
Gogus et al. 2017;
Cheng et al. 2020;
Mottez et al. 2020;
Wei et al. 2021).

These findings further confirm that the X-ray bursts from 
magnetars are likely to be generated by some SOC process, 
which can be explained by a possible magnetic reconnection 
scenario in magnetars
(Zhang et al. 2012).

The general statistical properties of the XRBs further 
indicate that they could belong to a self-organized critical 
system (e.g., starquakes), making them very similar to the 
earthquake phenomena
(Xie et al. 2024).

The recent discoveries of a remarkable glitch/anti-glitch 
accompanied by fast radio burst (FRB)-like bursts from the 
Galactic magnetar SGR J1935+2154 have revealed the physical 
connection between the two
(Gao and Wei 2024).

The overall temporal and spectral characteristics (e.g., 
duration, fluence, peak flux, peak count rate, and total 
counts) of X-ray bursts (XRBs) originating from magnetars 
have been extensively studied to confirm their 
self-organized criticality (SOC) behaviors
(Xiao et al.~2024a, 2024b).

\subsection{Blazars}

Blazars are very compact objects associated with
super-massive black holes in the center of 
active, giant elliptical galaxies. They represent
a subgroup of {\sl active galactic nuclei (AGNs)},
which emit a relativistic beam or jet that is
aligned with the line-of-sight direction to Earth.
Due to this particular geometry, blazars exhibit
highly variable and highly polarized emission in
radio and X-ray emission (Aschwanden et al.~2016b;
Aschwanden 2025a).

Blazars are a rather extreme class of radio-loud
AGNs, consisting of
BL Lac objects (BL Lacs) and flat-spectrum radio
quasars. Owing to relativistic Doppler boosting,
blazar emission is dominated by the nonthermal
emission coming from its jet. Its spectral energy
distribution extends from radio to $\gamma$-rays,
and shows two bumps. The low-energy bump is believed
to be the synchrotron radiation of relativistic
electrons, while the origin of the high-energy
bump is under debate (inverse-Compton scattering,
relativistic proton synchrotron radiation, or
synchrotron radiation of secondary particles
produced in proton-photon interaction), 
(Yan et al.~2018). 		   

Shock acceleration in BL Lac jets has been probed
through X-Ray polarimetry
(Tavecchio et al. 2020a),
as well as with anisotropic electron populations in BL Lac jets         
(Tavecchio and Sobacchi 2020).

Observations of blazars includes Mrk 421 observed with XMM-Newton
(Yan et al.~2018). 
Another observed dataset includes the six
blazars PKS B1222+216, PKS 1510-089, 3C 273, 3C 279, 3C 454.3,
and CTA 102 (Tavecchio et al.~2020b; Meyer et al.~2019).
The size distributions fit power laws or lognormal distributions
equally well, which is no surprise because a log-normal converges
to a power law function at large sizes. 
The blazer 3C 454.3 was observed with the Fermi telescope
(Peng et al.~2023a).
Recent statistical SOC studies on blazars, mostly focusing on 
3C 454.3 and Mrk 421, are described in 
Zhang et al.~(2018), 
Yan et al. (2018), 
Peng et al. (2023b), and 
Aschwanden and Gogus (2025).

\subsection{Black Hole Systems}

Sagittarius $A^*$ , the most prominent
black hole candidate in our galaxy,
is located at the center of the Milky Way. 
The routinely flaring events from Sagittarius $A^*$
trace dynamic, high-energy processes in the
immediate vicinity of the supermassive black hole
(Yuan et al. 2017).
A systematic study with Chandra of Sagittarius $A^*$
revealed the X-ray flare statistics in 
black holes. The flares
are detected from Sagittariusr $A^*$ during 1999-2012.
The fluence distribution is found to be
well-characterized by $\alpha_E=1.73\pm0.20$ 
(Yuan et al. 2017). 

A relatively small dataset of 38 events of
Sagittarius $A^*$ events, observed with Chandra
ACIS-S/HETGS, yielded a fluence distribution
with a slope of $\alpha_E=1.65\pm0.17$,
which agrees well with the FD-SOC model 
(Li et al.~2016).

Comparisons of the size distributions of the
tidal disruption event observed with Swift
J1644+57, the black hole system Sagittarius $A^*$ ,
and AGN M87 were found to be similar to
solar flares observed with RHESSI, and thus 
are explained with a magnetic reconnection
mechanism in the framework of a SOC model
(Wang et al.~2015).

The quiescent emission of Sagittarius $A^*$ has been 
explained in terms of the radiatively 
inefficient inflow/outflow model 
(Yuan et al.~2018).

\section{Conclusions}

In this review we summarize progress that has been made
regarding self-organized criticality (SOC) models and
the interpretation of
related observations during the last decade (2015-2025).
This time range complements a previous review covering
the years (1985-2015), (Aschwanden et al.~2016b).
The first three decades deal with the initial SOC 
discoveries (which are not repeated here for brevity
reasons), while the fourth decade deals with more 
advanced and refined SOC modeling, as well as
highlights new trends. The selected SOC
phenomena are limited to astrophysical observations,
such as solar flares, stellar flares, pulsar glitches,
soft gamma-ray bursts (SGR), fast radio bursts (FRB), 
and the black-hole candidate Sagittarius $A^*$.
We summarize our findings in the following.

\begin{enumerate}
\item{{\bf Ideal power law distribution function:}
	Traditional treatments of the size 
	distribution function of SOC avalanches
	assume an ``ideal'' or ``straight''
	size ($s$) distribution function 
	within a fitting range of
	$s_{\rm min} \le s \le {\rm max}$.
	The logarithmic size distribution,
	$N(s) \propto s^{-\alpha}$, contains
	a linear relationship between the 
	logarithmic variables, $\log(N) \propto \log(s)$.
	The original definition of SOC models 
	(Bak et al.~1987) emphasizes: 
	{\sl Power laws are the hallmarks of 
	self-organized criticality}.
	In reality, however, deviations from
	ideal power law functions and their
	modeling attempts became equally important
	during the last decade. Nevertheless,
	size distributions with ideal power laws ranges
	are frequently found in astrophysical
	phenomena, especially in datasets with
	large-number statistics, such as in
	solar flares (Aschwanden et al.~2000; 
	Lu et al.~1993; Shimizu 1995; 
	Crosby et al.~1993), and stellar flares
	(Aschwanden and G\"udel 2021), as shown 
	in form of a sythetic size distribution
	in Fig.~(6).}

\item{{\bf Thresholded size distribution:}
	Ideally, the power law function should
	exhibit a sharp cutoff at the smallest
	observed size $s_{\rm min} \le s$, but can be
	arbitraily small (but finite, in order
	to avoid the singularity at $s=0$). 
	The limiting value $s_{\rm min}$ depends 
	on the instrumental sensivity 
	of the observing instrument. In practice,
	the instrumental sensitivity rolls over 
	gradually into a flattened distribution,
	see Fig.(2). Practically, this part of
	the size distribution can be modeled,
	for instance with a Pareto Type II
	function, i.e., $N(s) \propto [s_0+s]^{-\alpha}$, 
 	marked with PM in Fig.~(2). While an
	implementation of the PM model can
	constrain a more accurate power law fit,
	we should be aware that a Pareto function
	is based on an {\sl ad hoc} assumption and
	thus does not reveal any new physics of SOC
	avalanches. Examples of thresholded size
	distributions can be found in Fig.~(1)
	which manifest incomplete sampling
	on the left side of the size distribution 
	(Aschwanden et al.~2015).}
		
\item{{\bf Background subtraction:}
	Characterizing the flux $f(t)$ in a SOC avalanche
	time profile, there is always an event-unrelated
	background profile $f_{\rm BG}(t)$ that needs to be
	subtracted before SOC avalanches are detected in
	the time profile $F(t)=f(t)+f_{\rm BG}(t)$. For instance,
	solar flares may dominate the soft X-ray time profile $f(t)$, 
	while (co-spatial) galactic soft X-ray emission may dominate 
	the soft X-ray source background emission $f_{\rm BG}(t)$.
	This correction is particulary important in
	small avalanche sizes, where the power law slope
	steepens after background subtraction. Examples
	are shown in Li et al.~(2016)}.\
	 
\item{{\bf Finite system size effects:}
	The scale-free relationship $N(s) \propto s^{-\alpha}$
	holds for a limited range,
	$s_{\rm min} \le s \le s_{\rm max}$, but brakes down at
	the maximum value $s=s_{\rm max}$, because every physical
	system is finite. For solar or stellar flares, for instance,
	an upper limit is given by the maximum size of active regions,
	sunspots, or starspots, ultimately limited by the solar or stellar
	diameter. The size distribution at the upper end has generally
	been modeled with an exponentially falling-off function,
	$N(s) \propto {\exp(-s)}$, 
	which is an empirical approximation at best,
	but lacks a physical model, strictly speaking. A combined
	size distribution that includes incomplete sampling (on the
	left-hand side) and finite-system size effecs 
	(on the right-hand side) is shown in Fig.~(2). Inbetween
	these two boundaries, the size distribution of Kepler-observed
	stellar flares reveals a power law extending over 
	4 orders of magnitude (Fig.~2), which can be stretched
	to 13 orders of magnitude (Fig.~6) in a synthesized size
	distribution (Aschwanden and Schrijver (2025).}

\item{{\bf Power law slope:}
	The most interesting parameter in SOC statistics is the
	power law slope $\alpha_s$, because it constrains the
	physical scaling laws, in particular the fractal
	dimension, the flux-volume scaling, classical
	diffusion, and the scale-freeness. From the theoretical
	fractal-diffusive FD-SOC model we derived power law
	slopes for the fluxes, $\alpha_F=1.80$, and for the
	fluences or energies, $\alpha_E=1.67$, which mostly
	agree with the observed size distributions (Table 3),
	while the misfits can be arguably explained with
	small-number statistics, thresholded (incomplete)
	sampling, and finite system size effects. As a 
	rule-of-thumb, at least 2 orders of magnitude
	are required to obtain a reliable value for the
	power law slope.} 

\item{{\bf Waiting time distribution:} Sampling of time-ordered
	sizes yields an exponential distribution function in the
	case of random (Poission) statistics on one side, and a
	power law distribution function in the case of
	nonstationary event rates on the other side (Fig.~3).
	The diagnostic value of power law slopes obtained from
	waiting time distributions requires the knowledge of
	the event rate $\lambda(t)$ (Eq.~4). 
	Earlier interpretations of
	the waiting time power law function ignored the
	effect of the variability of the event rate, and
	therefore were not conclusive, while modern studies
	take this variability into account
	(Aschwanden 2019c; Aschwanden and Johnson 2021;
	Aschwanden et al~2021; Nurhan et al.~2021;
	Aschwanden and Dudok de Wit 2021).}

\item{{\bf Cellular automaton:}
	The original approach of cellular automaton
	mechanisms to describe the spatio-temporal
	dynamics of SOC processes is based on
	next-neighbor interactions in an Euclidean
	d-dimensional lattice grid (d=1,2,3), which 
	requires numerical computer simlations. However,
	although such cellular automaton simulations
	yielded size distributions with power law 
	functions, some even achieving values 
	similar to those of observed astrophysical
	phenomena, it turned out that the BTW-type 
	numerical simulations are not necessary 
	to explain SOC in general,
	because the same power law slopes could be
	accomplished with simple scaling laws with 
	universal validity, which represents a
	paradigm shift of the SOC phenomenon.}

\item{{\bf Paradigm shift:}
	The Bak-Tang-Wiesenfeld (BTW) model, which
	simulates SOC processes, defines the SOC model 
	by means of a cellular automaton method, where 
	each avalanche in a SOC system evolves like the
	domino effect in a chain reaction.
	The FD-SOC model, in contrast, makes use of
	universal scaling laws, which can be applied
	with simple analytical expressions,
	and this way represents a paradigm shift.
	In essence, the BTW model uses 
	next-neighbor interations to describe the
	spatial spread of a SOC avalanche, while
	the FD-SOC model uses classical diffusion
	to describe the spatio-temporal expansion
	of a SOC avalanche.}

\end{enumerate}

We see that new trends in the SOC literature 
focus on quantitative modeling of physical systems,
while the initial studies mostly reported on
SOC discoveries in a qualitative way.  
We cannot claim that we understand the SOC
phenomenon unless the predictions of a
theoretical SOC model match the observations 
within the statistical uncertainties.  
A synopsis of the observed power law slopes 
and their uncertainties are listed in Table 3,
which may be substantially improved in future,
using larger event statistics and proper 
treatment of background subtraction, incomplete 
sampling, and finite system size effects.
Interestingly, we do not need to understand
the physical processes that produce a SOC
avalanche, such as magnetic reconnection
in the case of solar or stellar flares,
(see Fig.~4 for a diagram of possibly
involved physical processes).
because the power law slope of each
SOC parameter can be predicted from the
FD-SOC model, which fulfills the 
universality condition of SOC processes.
The only physics that is needed to explain
SOC is the flux-volume correlation, 
the classical diffusion, the fractal dimension,
and the scale-freeness, which all appear to
be universal.

\acknowledgements
This work was stimulated by the organizers of a 
workshop on “Mechanisms for extreme event generation” 
(MEEG) at the Lorentz Center at Snellius, Leiden, 
The Netherlands, July 8-12, 2019, organized by Drs. 
Norma Bock Crosby, Bertrand Groslambert, Alexander 
Milovanov, Jens Juul Rasmussen, and Didier Sornette. 
The author acknowledges the hospitality and partial 
support of two previous workshops on “Self-Organized 
Criticality and Turbulence” at the Inter national 
Space Science Institute (ISSI) at Bern, Switzerland, 
during October 15-19, 2012, and September 16-20, 2013,
as well as constructive and stimulating discussions 
with Sandra Chapman, Paul Charbonneau, Aaron Clauset, 
Norma Crosby, Michaila Dimitropoulou, Manolis 
Georgoulis, Stefan Hergarten, Henrik Jeldtoft Jensen, 
James McAteer, Shin Mineshige, Laura Morales, Mark 
Newman, Naoto Nishizuka, Gunnar Pruessner, John Rundle, 
Carolus Schrijver, Surja Sharma, Antoine Strugarek, 
Vadim Uritsky, and Nick Watkins. This work was partially 
supported by NASA contracts NNX11A099G “Self-organized 
criticality in solar physics”, NNG04EA00C of the SDO/AIA 
instrument, and NNG09FA40C of the IRIS instrument.

\clearpage

\begin{table}
\begin{center}
\caption{Theoretically predicted power law slopes $\alpha_x$ of the
standard FD-SOC (fractal-diffusive self-organized crticiality) model 
with dimensionality $d=3$ (Aschwanden 2015a).}
\medskip
\begin{tabular}{llll}
\hline
Parameter                    &Generalized                  & FD-SOC     & Power law   \\
                             &Power law                    & Power law  & observation \\
                             &Slope                        & Slope      & $\alpha_x$   \\
\hline
Euclidean Dimension          &$d$                          & (3/1)=3.00 & $3.3\pm0.3$ \\
Diffusion type               &$\beta$                      & (1/1)=1.00 &             \\
Fractal dimension            &$D_d=d-(1/2)=$               & (5/2)=2.50 &             \\
Length                       &$\alpha_L=d=$                & (3/1)=3.00 & $3.3\pm0.3$ \\
Area                         &$\alpha_A=1+(d-1)/D_A=$      & (7/3)=2.33 & $2.1\pm0.3$ \\
Volume                       &$\alpha_V=1+(d-1)/D_V=$      & (9/5)=1.80 & $1.7\pm0.2$ \\
Duration                     &$\alpha_T=1+(d-1) \beta/2=$  & (2/1)=2.00 & $2.1\pm0.2$ \\
Flux                         &$\alpha_F=1+(d-1)/(D_V)=$    & (9/5)=1.80 & $2.2\pm0.2$ \\
Energy                       &$\alpha_E=1+(d-1)/(d  )=$    & (5/3)=1.67 & $1.8\pm0.2$ \\
\hline
\end{tabular}
\end{center}
\end{table}

\begin{table}
\begin{center}
\caption{Energy ranges for solar and stellar flares.}
\normalsize
\medskip
\begin{tabular}{ll}
\hline
Energy range        & Wavelength range                  \\
(erg)               &                                   \\
\hline
$E=10^{24}-10^{27}$ & EUV solar nanoflares              \\
$E=10^{27}-10^{29}$ & Soft X-ray solar microflares      \\
$E=10^{29}-10^{32}$ & Hard X-ray solar flares           \\
$E=10^{32}-10^{36}$ & White-light stellar flares        \\
$E=10^{36}-10^{37}$ & White-light stellar superflares   \\
\hline
\end{tabular}
\end{center}
\end{table}

\clearpage

\begin{table}
\begin{center}
\caption{Synthesis of power law slopes in  astrophysical SOC phenomena.}
\normalsize
\medskip
\begin{tabular}{lllll}
\hline
Power law       & Power law	& Instrument & Phenomenon   & Reference \\
slope		& slope         &            &              &           \\
flux		& energy        &            &              &           \\
$\alpha_F$      & $\alpha_E$    &            &              &           \\
\hline                           
...             & 1.80$\pm$0.2 	& AIA/SDO& solar flares          & A15a \\
...             & 1.3-1.7       & RHESSI,Fermi & TGF             & O12  \\
1.80$\pm$0.19   & ...           & UBF            & stellar flares        & AG21 \\  
2.08$\pm$0.37   & ...           & XUV            & stellar flares        & AG21 \\ 
1.93$\pm$0.40   & ...           & KEPLER         & stellar flares        & AG21 \\  
2.16$\pm$0.10   & ...           & TESS           & stellar flares        & AG21,T20 \\  
2.10$\pm$0.21   & ...           & KEPLER         & stellar flares G-type & AG21 \\ 
2.36$\pm$0.25   & ...           & KEPLER         & stellar flares F-type & AG21 \\ 
1.89$\pm$0.35   & ...           & KEPLER         & stellar flares M-type & AG21 \\  
2.02$\pm$0.33   & ...           & KEPLER         & stellar flares K-type & AG21 \\  
1.12$\pm$0.08   & ...           & KEPLER         & stellar flares A-type & AG21 \\  
1.90$\pm$0.10   & ...           & KEPLER         & stellar flares Giants & AG21 \\  
2.26$\pm$0.11   & ...           & Fermi          & SGR J1550+5418        & X24d \\ 
2.05$\pm$0.07   & ...           & Fermi          & SGR J1935+2154        & X24d  \\
1.2-1.9         & ...           & XMM-Newton     & SFXT                  & Z22  \\
...             & 1.7           & ASKAP          & FRB                   & L19  \\
1.8$\pm$0.3     & ...           & Arecibo        & FRB 121102            & G19  \\
1.86$\pm$0.02   & ...           & FAST           & FRB                   & Z21  \\
1.70$\pm$0.03   & ...           & FAST           & FRB                   & Z21  \\
2.60$\pm$0.15   & ...           & FAST           & FRB (early bursts)    & Z21  \\
1.57-1.63       & ...           & Swift          & FRB                   & W23  \\
...             & 1.73$\pm$0.20 & Chandra        & Sagittarius $A^*$     & Y17  \\ 
...             & 1.65$\pm$0.17 & Chandra        & Sagittarius $A^*$     & L15  \\ 
\hline   
1.80		& 1.67		& Theory	 & FD-SOC model & Section 2.4 \\
\hline
\end{tabular}
\end{center}
References: A15=Aschwanden et al.~2015a; 
            O12=Ostgaard et al.~2012; AG21=Aschwanden and G\"udel 2021;
	    T20=Tu et al.~2020; X24=Xiao et al.~2024; Z22=Zhang et al.~2022;
	    L19=Lu and Piro 2019; G19=Gourdji et al.~2019;
	    Z21=Zhang et al.~2021; W23=Wang et al.~2023; Y17=Yuan et al.~2017;
	    L15=Li et al.~2015; 
	    FRB=Fast Transient Radio bursts; 
	    SFXT=Supergiant fast X-ray transients;
\end{table}

\clearpage

\def\ref#1{\par\noindent\hangindent1cm {#1}} 
\section*{References}

\ref{Abramenko, V.I.
	2015, Geomagnetism and Aeronomy 55, 860-865}
	{\sl Possibilities of predicting flare productivity
	based on magnetic field power spectra in active regions}
\ref{Abramenko, V.I. and Yurchyshyn, V.B.
	2020, MNRAS 497, 5405-5412.}
	{\sl Analysis of quiet-Sun turbulence on the basis of
	SDO/HMI and Goode solar telescope data}
\ref{Abramenko, V.I.
	2024, Geomagnetism and Aeronomy 60/7, pp.801-803.}
	{\sl Self-organized criticality of solar magnetism}
\ref{Abramenko, V.I. and Suleymanova, R.
	2024, Sol.Phys. 299/3, id.31.} 
	{\sl Correlation functions of photospheric magnetic
	fields in solar active regions}
\ref{Alipour, N. and Safari, H.
	2015, ApJ 807/2, id.175, 9 pp.}
	{\sl Statistical properties of solar coronal bright
	points}
\ref{Alipour, N., Safari, H., Verbeeck, C., Berghmans, D. et al. 
	2022, AA 663, id.A128, 12.}
	{\sl Automatic detection of small-scale EUV brightenings
	observed by the Solar Orbiter/EUI} 
\ref{Antolin,A., Pagano, P., Testa, P., Petralia, A. et al. 
	2021, Nature Astronomy, 5, p.54-62.} 
	{\sl Reconnection nanojets in the solar corona}
\ref{Aschwanden, M.J., Tarbell, T.D., Nightingale, W.,
        Schrijver, C.J., Title A., Kankelborg, C.C., and Martens, P. 2000,
        {\sl Time variability of the ``quiet'' Sun observed with TRACE.
        II. Physical parameters temperature evolution, and energetics
        of extreme-ultraviolet nanoflares}, ApJ 535, 1047-1065.}200
\ref{Aschwanden, M.J. and McTiernan, J.M. 
	2010, ApJ 717, 683-692.}
        {\sl Reconciliation of waiting time statistics of solar flares
        observed in hard X-rays},
\ref{Aschwanden, M.J.
	2011, Self-organized criticality in astrophysics. The
	statistics of nonlinear processes in the universe,
	Cambridge University Press, Cambridge.}
\ref{Aschwanden, M.J. 
	2014, ApJ 782, 54.}
        {\sl A macroscopic description of self-organized systems and
        astrophysical applications}  
\ref{Aschwanden, M.J., Xu, Y., Jing, J. 
	2014a, ApJ 797:50, (35pp).}
	{\sl Global energetics of solar flares: I. Magnetic energies}
\ref{Aschwanden, M.J., Sun, X., and Liu, Y.
	2014b, ApJ 785:34 (27pp.)}
	{\sl The magnetic field of active region 11158 during the
	February 12-17 flares: Differences between photospheric
	extrapolation and coronal forward-fitting methods}
\ref{Aschwanden, M.J. 
	2015, ApJ 814, 19 (25pp).}
 	{\sl Thresholded powerlaw size distributions of instabilities 
	in astrophysics}
\ref{Aschwanden, M.J. Boerner, P., Ryan, D., Caspi, A., et al. 
	2015a, ApJ 802:53 (20pp).}
 	{\sl Global energetics of solar flares: II. Thermal Energies} 
\ref{Aschwanden, M.J., Schrijver, C.J., and Malanushenko, A.
	2015b, Sol.Phus 290/10, pp.2765-2789.}
	{\sl Blind stereoscopy of the coronal magnetic field}
\ref{Aschwanden, M.J.
	2016a, ApJ 831:105, (34pp).}
	{\sl Global energetics of solar flares. IV. Coronal mass ejections
	energetics}
\ref{Aschwanden, M.J.
	2016b, ApJ 224/2, id.25, 32pp.}
	{\sl The vertical-current approximation nonlinear force-free
	field code - Description, performance tests, and measurements
	of magnetic energies dissipated in solar flares}
\ref{Aschwanden, M.J., Holman G., O'Flannagain, A., Caspi, A., et al. 
	2016a, ApJ 832:27, (20pp).}	
 	{\sl Global energetics of solar flares: III. Nonthermal Energies}
\ref{Aschwanden, M.J., Crosby, N., Dimitropoulou, M., Georgoulis, M.K., et al.
	2016b, SSRv 198, 47.} 
	{\sl 25 Years of Self-Organized Criticality: Solar and Astrophysics} 
\ref{Aschwanden, M.J., Reardon, K., at Jess, D.B. 
	2016c, ApJ 826, id 61, 18pp.}
	{\sl Tracing the chromospheric and coronal magnetic field with
	AIA, IRIS, IBIS, and ROSA}
\ref{Aschwanden, M.J.
	2017, ApJ 847:27, (19pp.}
	{\sl Global energetics of solar flares. VI. Refined energetics 
	of coronal mass ejections}
\ref{Aschwanden, M.J. and Peter, H. 2017,
        {\sl The width distribution of solar coronal loops and strands -
        Are we hitting rock bottom ?}
        ApJ 840:4 (24pp).}
\ref{Aschwanden, M.J., Caspi, A., Cohen, C.M.S., Holman, G. et al. 
	2017, ApJ 836, id 17, 17pp.}
	{\sl Global energetics of solar flares. V. Energy closure
	in flares and coronal mass ejections}
\ref{Aschwanden, M.J.
	2018, ApJ 866:73 (13pp).}
	{\sl Convection-driven generation of ubiquitous coronal waves}
\ref{Aschwanden, M.J. and Gopalswamy, N.
	2019, ApJ 877:149, (14pp).}
	{\sl Global energetics of solar flares. VII. Aerodynamic drag
	in coronal mass ejections}
\ref{Aschwanden, M.J. 
	2019a, ApJ 881:1 (22p).}
	{\sl Global energetics of solar flares. VIII. The low-energetic
	cutoff}
\ref{Aschwanden, M.J. 
	2019b ApJ 885: 49, (21 pp).}
	{\sl Global energetics of solar flares. IX. Refined magnetic
	modeling},
\ref{Aschwanden, M.J. 
	2019c, ApJ 887:57, (12pp).}
 	{\sl Non-stationary fast-driven self-organized criticality 
	in solar flares},
\ref{Aschwanden, M.J.
	2019d, ApJ 874, id.131, 10pp.}
	{\sl Helical twisting number and braiding linkage number of
	solar coronal loops}
\ref{Aschwanden, M.J.
	2019e, ApJ 880:105, 16pp.}
	{\sl Self-organized criticality in solar and stellar flares:
	Are extreme events scale-free ?}
\ref{Aschwanden, M.J. 
	2020a, ApJ 895:134 (11pp).}
	{\sl Global energetics of solar flares. X. Petschek reconnection
	rate and Alfv\'en Mach number of magnetic reconnection outflows}
\ref{Aschwanden, M.J.
	2020b, ApJ 897,16 (11pp).}
	{\sl Global energetics of solar flares. XI. Flare magnitude
	prediction of the GOES class}
\ref{Aschwanden, M.J.
	2020c, ApJ 903:23, (14pp).}
	{\sl Global energetics of solar flares. XII. Energy scaling laws}
\ref{Aschwanden, M.J.  
        2021, ApJ 909, 69.}
        {\sl Finite system-size effects in self-organizing criticality
        systems} 
\ref{Aschwanden, M.J. and Dudok de Wit, T. 2021,
	{\sl Correlation of the Sunspot Number and the Waiting-time 
	Distribution of Solar Flares, Coronal Mass Ejections, and 
	Solar Wind Switchback Events Observed with the Parker Solar Probe},
	ApJ 812, id.94, 11pp.}
\ref{Aschwanden, M.J. and G\"udel, M.
 	2021, ApJ 910:41.}
 	{\sl Self-organized criticality in stellar flares}
\ref{Aschwanden, M.J. and Johnson, J.R. 2021,
	{\sl The solar memory from hours to decades},
	ApJ 921, id.82.}
\ref{Aschwanden, M.J., Johnson,J.R., and Nurhan,Y. 2021,
 	{\sl The Poissonian origin of power laws in solar flare 
	waiting time distributions},
 	ApJ 921:166 (14pp).}
\ref{Aschwanden, M.J. 
	2022a, ApJ 934, 3.}
	{\sl Reconciling power law slopes in solar flare and
	nanoflare size distributions},
\ref{Aschwanden, M.J. 
	2022b, ApJ 934:33 (27pp).} 
        {\sl The fractality and size distributions of astrophysical
        self-organized criticality systems},
\ref{Aschwanden, M.J.
	2022c, ApJ (subm).}
	{\sl Global energetics of solar flares. VIII. The Neupert effect
	and acceleration of coronal mass ejections}
\ref{Aschwanden M.J. and Nhalil, N.V.
	2023a, Frontiers in Astron. and Space Sci. 10.999346.}
	{\sl The universality of power law slopes in the solar
	photosphere and transition region observed with HMI and IRIS}	
\ref{Aschwanden M.J. and Nhalil, N.V.
	2023b, Frontiers in Astron. and Space Sci. 9.999319.}
	{\sl Interface region imaging spectrograph (IRIS) observations
	of the fractal dimension in the solar atmosphere}
\ref{Aschwanden, M.J. 
	2025a,
        {\sl Power Laws in Astrophysics. Self-Organzed Criticality
        Systems}, Cambridge University Press: Cambridge.}
\ref{Aschwanden, M.J.
	2025b, ApJ 980:209 (7pp).} 
	{\sl Universal constants in self-organized riticality systems}
\ref{Aschwanden, M.J. and Gogus, E.
	2025, ApJ 978/1, id.19, 11pp.}
	{\sl Testing the Universality of Self-organized Criticality 
	in Galactic, Extragalactic, and Black Hole Systems}
\ref{Aschwanden, M.J. and Schrijver, C.J.
	2025, ApJ 987:140 (8pp).}
	{\sl Self-organized criticality across 13 orders of magnitude
	in the solar-stellar connection}
\ref{Bak, P., Tang, C., and Wiesenfeld, K. 
	1987, Physical Review Lett. 59(27), 381.}
        {\sl Self-organized criticality: An explanation of 1/f noise}
\ref{Balona, L.A.
	2015, MNRAS 447, 3, p.2714-2725.}
	{\sl flare stars across the H-R diagram}
\ref{Balona, L.A.
	2019, MNRAS 490, 2112-2116.}
	{\sl Evidence for spots on hot stars suggests major
	revision of stellar physics} 
\ref{Banerjee, A., Bej, A., Chatterjee, T.N., and Majumdar, A.
        2019, Earth.Spa.Sco. 6/4, pp.565-576.}
	{\sl A SOC approach to study the solar wind-magnetosphere energy
	coupling}
\ref{Berger, M.A., Asgari-Targhi, M., and DeLuca, E.E.
	2015, J. Plasma Physics 81, id.395810404.}
	{\sl Self-organized braiding in solar coronal loops}.
\ref{Bolzan, M.J.A.
	2018, Physica A. Stat.Mech.Appl. 503, p.1182-1188.}
	{\sl A modeling substorm dynamics of the magnetosphere using
	self-organized criticality approach}
\ref{Borucki, B.W., Koch, D., Basri, G., Batalha, N. et al.
	2010, Science, 327, Issue 5968, pp.977.}
	{\sl Kepler planet-detection mission: Introduction and
	first results}
\ref{Bottke, D., Vokrouhlicky, D. Minton, A., Nesvorny, R. et al.
	2012, Nature 485(7936), 78-81.}
	{\sl An Archean heavy bombardment from a destabilized
	extension of the asteroid belt}
\ref{Cadavid, A.C., Miralles, M.P., and Romich, K.
	2019, ApJ 886, 143, (12pp).}
	{\sl Comparison of the scaling properties of EUV
	intensity fluctuations in coronal hole and Quiet-Sun regions}
\ref{Candelaresi, S., Pontin D.I., Yeates, A.R., Bushby P.J.
	2018, ApJ 864:157 (9pp).}
	{\sl Estimating the rate of field line braiding in the
	solar corona by photospheric flows} 
\ref{Chatterjee, S., Law, C.J., Wharton, R.S., Burke-Spolaor, S. et al.
	2017, Nature, Volume 541, issue 7635, pp.58-61.}
	{\sl The direct localization of a fast radio burst and its host}
\ref{Chen, Y. and Hu Q. 2022, ApJ 924:43 (12pp).}
	{\sl Small-scale magnetic flux ropes and their properties
	based on in situ measurements from the Parker Solar Probe}
\ref{Cheng, Y., Zhang, G.Q., and Wang, F.Y.
	2020, MNRAS 491/1, 1498-1505.}
	{\sl Statistical properties of magnetar bursts and FRB 121102}
\ref{Chesny, D.L., Oluseyi, H.M., and Orange, N.B.
	2016, ApJ 822, id.72, 12pp.}
	{\sl Dynamic flaring non-potential fields on quiet
	Sun network scales}
\ref{Cliver, E.W., Ling, A.G., Belov, A., and Yashiro, S.
	2012, ApJL 756:L29 (4pp).}
	{\sl Size distributions of solar flares and solar energetic
	particle events}
\ref{Cliver, E.W. and D'Huys, E.
	2018, ApJ 864, id.48, 11pp.}
	{\sl Size distributions of solar proton events and their
	associated soft X-ray flares: Application of the maximum
	likelihood estimator}
\ref{Crosby, N.B., Aschwanden, M.J., and Dennis, B.R. 1993,
        {\sl Frequency distributions and correlations of solar flare
        parameters}, SoPh 143, 275-299.}
\ref{Daei, F., Safari, H., and Dadashi, N.
	2017, ApJ 845, id.36, (8pp).}
	{\sl Complex network for solar active regions}
\ref{Davenport, J.R.A.
	2016, ApJ 829:23 (12pp).}
	{\sl The KEPLER catalog of stellar flares}
\ref{Dudok de Wit, T., Krasnoselskikh, V.V., Bale, S., Bonnell, J.W. et al.
	2020, ApJSS 246:39 (10pp).}
	{\sl Switchbacks in the near-Sun magnetic field: Long memory and
	impact on the turbulence cascade}
\ref{Enoto, T., Shibata, S., Kilaguchi, T. et al.
	2017, ApJSS 231/1, id.8, 21 pp.}
	{\sl Magnetar Broadband X-Ray Spectra Correlated with Magnetic 
	Fields: Suzaku Archive of SGRs and AXPs Combined with 
	NuSTAR, Swift, and RXTE.}
\ref{Farhang, N., Safari, H., and Wheatland, M.S. 
	2018, ApJ 859, id.41, 10pp.}
	{\sl Principle of minimum energy in magnetic reconnection
	is a self-organized critical model for solar flares}
\ref{Farhang, N., Wheatland, M.S., and Safari, H. 
	2019, ApJL 883:L20 (7pp).}
	{\sl Energy balance in avalanche models for solar flares}
\ref{Fatuzzo, F., Adams, F.C., Feinstein, A.D., and Seligman, D.Z.
	2023, ApJ 954, id.15, 10pp.}
	{\sl Avalalanches and the distribution of reconnection
	events in magnetized circumstellar disks}
\ref{Feinstein, A.D., Seligman, D.Z., G\"unther, M.N., and Adams, F.C.
	2022, ApJL 925:L9 (6pp).}
	{\sl Testing self-organized criticality across the main
	sequence using stellar flares} 
\ref{Fishman et al. 
	1994, Science 264, 1313-1316.}
	{\sl Discovery of intense gamma-ray flashes of atmospheric
	origin}
\ref{Fox, N.J., Velli, M.C., Bale, S.D. et al.
	2016, SSRv 204, 7.}
	{\sl The Solar Probe Plus Mission: Humanity's first visit
	to our star}
\ref{Gao, C.Y. and Wei, J.J.
	2024, ApJ 968:40 (8pp.)}
	{\sl A comparative analysis of scale-invariant phenomena
	in repeating fast radio bursts and glitching pulsars}
\ref{Georgoulis, M.K., Nindos, A., Zhang, H.
	2019, Phil.Trans.Royal Society A 377, 2184, p.20180094.}
	{\sl The source and engine of coronal mass ejections}
\ref{Gheibi, A., Safari, H., and Javaherian, M.
	2017, ApJ 847, id.115, (12 pp).
	{\sl The solar flare complex network}
\ref{Giannattasio, F. and Consolini, G.
	2021, ApJ 906/2, id.142, 7pp.}
	{\sl The complex nature of magnetic element transport in the
	Quiet-Sun multiscaling character}
\ref{Giannattasio, F., Consolini, G., Berilli, F., and DelMoro, D.
	2019, ApJ 878/1, id33, 9pp.}
	{\sl The complex nature of magnetic element transport in
	the Quiet-Sun: The \'evy-walk character}
\ref{Giannattasio, F., Consolini, G., Berilli, F., and DelMoro, D.
	2020, ApJ 904/1, id.7, 7pp.}
	{\sl Magnetic energy balance in the Quiet-Sun on
	supergranular spatial and temporal scales}
\ref{Giorgi, F., Ermolli, I., Romano, P., Stangalini, M. et al.
	2015, Sol.Phys. 290/2, pp.507-525.}
	The signature of flare activity in multifractal measurements
	of active regions observed by SDO/HMI}
\ref{Gizis, J.E., Paudel, R.R., Mullan, D., Schmidt, S.J. et al.
	2017a, ApJ 838:22 (6pp).}
	{\sl K2 Ultracool dwarfs survey. I. Photometry of an L
	dwarf superflare}
\ref{Gizis, J.E., Paudel, R.R., Mullan, D., Schmidt, S.J. et al.
	2017b, ApJ 845:33 (12pp).}
	{\sl K2 Ultracool dwarfs survey. II. The white light flare
	rate of young brown dwarfs}
\ref{Gogus, E., Lin, L., Roberts, O.J., Chakraborty, M. et al.
	2017, ApJ 835/1, id.68, 8pp.}
	{\sl Burst and Outburst Characteristics of Magnetar 4U 0142+61}
\ref{Golovko, A.A. and Sladkhutidinova, I.I. 
	2015, Astronomy Reports 59/8, pp.776-790.}
	{\sl Evolution of solar active regions: Detecting the emergence
	of new magnetic field through multifractal segmentation}
\ref{Golovko, A.A. and Sladkhutidinova, I.I. 
	2018, J.Atmos.Solar-Terr.Phys. 179, p.120-127.}
	{\sl Detecting the solar new magnetic flux regions on
	the base of vector magnetograms}
\ref{Gourdji, K., Michilli, D., Spitler, L.G., Hessels, J.W.T, et al.
	2019, ApJ 877, 2, L19, 12pp.}
	{\sl A sample of low-energy bursts from FRB 121102}
\ref{Guerreiro, N., Haberreiter, M., Hansteen, V., and Schmutz, W. 
	2015, ApJ 813, id.61, 11pp.}
	{\sl Small-scale heating events in the solar atmosphere.
	I. Identification, selection, and implications for coronal
	heating}
\ref{Guerreiro, N., Haberreiter, M., Hansteen, V., and Schmutz, W. 
	2017, AA 603, A103.} 
	{\sl Small-scale heating events in the solar atmosphere.
	II. Lifetime, total energy, and magnetic properties}
\ref{Guidorzi, C., Dichiara, S., Frontera, F., Margutti, R. et al.
	2015, ApJ 801:57 (11pp).}
	{\sl A common stochastic rpocess rules gamma-ray burst
	prompt emission}
\ref{Guite, L.S., Charbonneau, P., and Strugarek, A. 
	2025a, Sol.Phys. 300, 82.}
	{\sl Avalanching together: A model for sympathetic flaring}
\ref{Guite, L.S., Strugarek, A., and Charbonneau, P. 
	2025b, AA 694, 74.}
	{\sl Flaring together. A preferred angular separation
	between sympathetic flares on the Sun}
\ref{Harko, T., Mocanu, G., and Stroia, N.
	2015, Astrophys.Spac.Sci. 357, id.84, 9pp.}
	{\sl Self-organized criticality in an one dimensional
	magnetized grid. Application to GRB X-ay afterglows}
\ref{Hermsen, W., Kuiper, L., Basu, R., Hessels, J.W.T. et al.
	2018, MNRAS 480, 3, p.3655-3670.}
        {\sl Discovery of synchonous X-ray and radio moding of
	PSR B0823+26}
\ref{Hood, A.W., Cargill, P.J., Browning, P.K. et al.
	2016, ApJ 817, id.5, 7pp.}
	{\sl An MHD avalanche in a multi-threaded coronal loop}
\ref{Hudson, H.S.
	1978 Sol.Phys. 57, pp237-240.}
	{\sl Threshold effect in second-stage acceleration}
\ref{Hudson, H.S. 
	2020, MNRAS 491/3, pp.4435-4441.}
	{\sl A correlation in the waiting-time distribution of
	solar flares}
\ref{Huppenkothen, D., Brewer, B.J., Hogg, D.W., Murray, I. et al.
	2015, ApJ 810/1, id.66, 21pp.}
	{\sl Dissecting Magnetar Variability with Bayesian 
	Hierarchical Models} 
\ref{Javaherian, M., Safari, H. Dadashi, N., and Aschwanden, M.J.
	2017, Sol.Phys. 292, 164.}
	{\sl Statistical properties of photospheric magnetic
	elements observed by the helioseismic and magnetic
	imager onboard the solar dynamics observatory}
\ref{Kahler, S.W.
	2013, ApJ 769:35 (5pp).}
	{\sl Does a scaling law exist between solar energetic
	particle events and solar flares}
\ref{Kanella, C., and Gudiksen, B.V.
	2017, AA 503, id.A83, 7pp.}
	{\sl Identification of coronal heating events in 3-D simulations}
\ref{Kawai, T. and Imada, S. 
	2022, ApJ 931/2, id.113, 10pp.} 
	{\sl Factors that determine the power-law index of an
	energy distribution of solar flares}
\ref{Kennedy, M.R., Clark, C.J., Voisin, G., and Breton, R.P.
	2018, MNRAS 477, 1, p.1120-1132.}
	{\sl Kepler K2 observations of the transitional
	millisecond pulsar PSR J1023+0038}
\ref{Knizhnik, K.J., Uritsky, V.M., Klimchuk, J.A. et al. 
	2018, ApJ 853, id.82, 14pp.} 
	{\sl Power-law statistics of driven reconnection in the
	magnetically closed corona}
\ref{Kouveliotou, C., Strohmayer, T., Hurley, K., van Paradijs, J. et al.
	1999, ApJ 510/2, pp.L115-L118.}
	{\sl Discovery of a magnetar associated with the soft gamma
	repeater SGR 1900+14}
\ref{Kuznetsov, A.A. and Kolotkov, D.Y.
	2021, ApJ 912:81 (16pp).}
	{\sl Stellar superflares observed simultaneously with Kepler
	and XMM-Newton}
\ref{Lamarre, H., Charbonneau, P., Strugarek, A. 
	2024, SoPh 299, id.13.}	
	{\sl Energy definition and minimization in avalanche models
	for solar rlares}
\ref{Lamy, P.L., Floyd, O., Boclet, B., Wojak, J. et al.
	2019, SSRv 215/5, id.39, 129 pp.}
	{\sl Coronal mass ejections over solar cycle23 and 24}
\ref{Lei, W.H., Li, C., Chen,F. et al.
	2020, MNRAS 494, pp.975-982.}
	{\sl Do the solar flares originating from an individual active
	region follow a random process or a memory-dependent correlation}'
\ref{Li, C., Yang, J., Li, C., Tian, Y. et al.
	2016, Scientia Sinica Physica, Mechanica and Astronomica 46/2, p.029501.}
	{\sl Self-organized criticality of the solar eruptions during solar cycle}
\ref{Li, X.J. and Yang, Y.P. 
	2023, ApJ 955/2, id.L34.}
	{\sl Signatures of the self-organized criticality phenomena in
	precursors of gamma-ray bursts}
\ref{Li, X.J., Liu J.M., Cheng, M. et al.
	2024, PASP 136/8, id.084204, 8pp.}
	{\sl Scale-invariant features of X-ray bursts from SGR J1935+2154
	detected by Insight-HXMT}
\ref{Li, X.J., Zhang, W.L., Yi, S.X., et al.
	2023, ApJSS 265/2, id.56, 14pp.}
	{\sl Evidence for the self-orgznized criticality phenomena
	in the prompt phase of shoft gamma-ray bursts}
\ref{Li, Y.P., Li, F., Zhang, P., Liu, S.M., et al.
	2016, Res.Astron.Astrophys. 16/10, id.161.}
	{\sl On the power-law distributions of X-ray fluxes from
	solar flares observed with GOES}
\ref{Li, C., Zhong, S.J., Xu, Z.G., He, H. et al.
	2018a, MNRAS 479, L138-L142.}
	{\sl Waiting time distributions of solar and stellar flares:
        Poisson process or with memory?}
\ref{Li, Y.P., Yuan, F., Wang, O.D., Chen, P.F.
	2018b, ApJ 810/1, id.19, 8pp.}
	{\sl Statistics of X-Ray Flares of Sagittarius $A^*$ : 
	Evidence for Solar-like Self-organized Criticality Phenomena}	
\ref{Love, J.J., Hayakawa, H., and Cliver, E.W.
	2020, Space Weather 17/1, pp.37-45.}
	{\sl On the intensity of the magnetic superstorm of September 1909}
\ref{Lu, W. and Piro, A.L.
	2019, ApJ 883/1, id.40, 8pp.}
	{\sl Implications from ASKAP fast radio burst statistics}
\ref{Lu, E.T., Hamilton, R.J., McTiernan, J.M., and Bromund, K.R. 1993,
        {\sl Solar flares and avalanches in driven dissipative systems},
        ApJ 412, 841-852.}
\ref{Lurie, J.C., Davenport, R.A., Hawley, S.L., Wilkinson, T.D. et al.
	2015, ApJ 800:95 (14pp).}
	{\sl KEPLER flares III: Stellar activity on GJ 1245A and B}
\ref{Lyu, F., Li, Y.P., Hou, S.J., Wei, J.Y. et al. 
	2020, Frontiers of Physics, 16/1, id.14501.}
	{\sl Self-organized criticality in multi-pulse gamma-ray bursts}
\ref{Lyu, F., Li, Y.P., Hou, S.J., Wei, J.Y. et al. 
	2021, Frontiers of Physics, 16/1, id.14501.}
	{\sl Self-organized criticality in multi-pulse gamma-ray bursts}
\ref{Maccary, R., Guidorzi, C., Amati, L., et al.
	2024, ApJ 965, 72.}
	{\sl Distributions of Energy, Luminosity, Duration, and Waiting 
	Times of Gamma-Ray Burst Pulses with Known Redshift Detected by 
	Fermi/GBM} 
\ref{Maehara, H., Shibayama, T., and Notsu, Y.
	2015, Earth, Planet and Space 67, id.59, 10pp.}
	{\sl statistical properties of superflares on solar-type stars
	based on 1-min cadence data}
\ref{Maehara, H., Notsu, S., Namekata, K., Honda, S. et al.
	{\sl 2017, PASJ 69, 3, 41.}
	{\sl Starspot activity and superflares on solar-type stars}
\ref{Mandelbrot, B.B. 1977,
        {\sl Fractals: form, chance, and dimension}, Translation of
        {\sl Les objects fractals}, W.H. Freeman, San Francisco.}
\ref{McAteer, R.T.J., Aschwanden,M.J., Dimitropoulou,M., Georgoulis, M.K. et al.
	2016, SSRv 198, 217.}
        {\sl 25 Years of Self-Organized Criticality: Numerical Detection
        Methods} 
\ref{Melatos, A., Gouglass, J.A., and Simula, T.P.
	2015, ApJ 807, 2, 132, 12 pp.}
	{\sl Persistent gravitational radiation from glitching pulsars}
\ref{Melatos, A., Dunn, L.M., Suvorova, S., Moran, W. and Evans, R.J.,
	2020, ApJ 896, 1, 78, 27 pp.}
	Pulsar glitch detection with a hidden Markov model} 
\ref{Meyer, M., Scargle, J.D., and Blandford, R.D.
	2019, ApJ 877/1, id.39, 30pp.}
	{\sl Characterizing the Gamma-Ray Variability of the 
	Brightest Flat Spectrum Radio Quasars Observed with the Fermi LAT}
\ref{Morina, D., Serra, I., Pulg, P., and Corral, A.
	2019, Scientific Reports 8, id.2393.}
	{\sl Probability estimation of a Carrington-like geomagnetic storm}
\ref{Mottez, F., Zarka, P., and Voisin, G.
	2020, AA 644,id.A145, 14pp.}
	{\sl Repeating fast radio bursts caused by small bodies 
	orbiting a pulsar or a magnetar} 
\ref{Newman, M.E.J. 
	2005, Contemporary Physics, Vol. 46, issue 5, pp.323-351.}
        {\sl Power laws, Pareto distributions and Zipf's law}
\ref{Notsu, Y., Honda, S., Maehara, Y. and
	2015a, PASJ 67, 3, id.32, 24pp.}
	{\sl High dispersion specroscopy of solar-type superflare stars.
	I. Temperature, surface gravity, metallicity and vsin i}
\ref{Notsu, Y., Honda, S., Maehara, Y. and
	2015b, PASJ 67, 3, id.33, 14pp.}
	{\sl High dispersion specroscopy of solar-type superflare stars.
	II. Stellar rotation starspots, and chromospheric activities}
\ref{Nurhan, Y., Johnson, J.R., Homan, J.R., Wing, S. and Aschwanden, M.J.
	2021, GRL 48, 16, ke94348.}
	{\sl Role of the solar minimum in the waiting time distribution
	throughout the heliosphere}
\ref{Ostgaard, N., Gjesteland, T., Hansen, R.S., Collier, A.B. et al.
	2012, JGR 117, A3, A03327.}
	{\sl The true fluence distribution of terrestrial gamma
	flashes at satellite altitude}
\ref{Peng, F.K., Wang, F.Y., Shu, X.W, Hou, S.J.
	2023a, MNRAS 518/3, pp.3959-3965.}
	{\sl Self-organized criticality in solar GeV flares}
\ref{Peng, F.K., Wei, J.J., and Wang, H.Q.
	2023b, ApJ 959/2, id.109, 8pp.}
	{\sl Scale Invariance in Gamma-Ray Flares of the Sun and 3C 454.3}
\ref{Pietras, M., Falewicz, R., Siarkowski, M., Bicz, K. et al.
	2022, ApJ 935:143 (19pp).}
	{\sl Statistical analysis of stellar flares from the first
	three years of TESS observations}
\ref{Piran, T.A
	1992, ApJ 389, p.45.}
	{\sl The implications of the Compton (GRO) observations
	for cosmological gamma ray bursts}
\ref{Pruessner, J.G. 2012,
 	Self-organised criticality. Theory, models and 
	characterisation, Cambridge University Press: Cambridge.}
\ref{Pontin, D.I. and Hornig, G.
	2015, ApJ 805, id.47.}
	{\sl The structure of current layers and degree of field-line
	braiding in coronal loops}
\ref{Rincon, R.F., Roudier, T., Schekochihin, A.A., and Rieutord, M.
	2017, AA 599, id.A69, 19pp.}
	{\sl Supergranulation and multiscale flows in the solar
	photosphere. Global observations vs. a theory of anisotropic 
	turbulent convection}
\ref{Ryan, D.F., Dominique, M., Seaton, D., Stegen, K., et al.  
	2016, AA 592, A133.}
        {\sl Effects of flare definitions on the statistics of derived
        flare distributions}
\ref{Sang, Y. and Lin, H.N.
	2024, MNAS 533, 782-879.}
	{\sl Quantifying the randomness and scale invariance of the
	repeating fast radio bursts}
\ref{Scholz, P., Spitler,, L.G., Hessels, J.W.T., Chatterjee, S., et al.
	2016, ApJ 833, id.177, 17pp.}
	{\sl The repeating fast radio bursts FRB 121102: Multi-wavelength
	observations and additional bursts}
\ref{Shapoval, A., LeMouel, J.L., Shnirman, M., and Courtillot, V.
	2018, AA 618, id.A183, 13.}
	{\sl Observational evidence in favor of scale-free
	evolution of sunspot groups}
\ref{Sharma, A.S., Aschwanden,M.J., Crosby,N.B., Klimas,A.J. 
	2016, SSRv 198, 167.}
        {\sl 25 Years of Self-Organized Criticality: Space and Laboratory
        Plamsas} 
\ref{Shimizu, T. 1995, {\sl Energetics and occurrence rate of active region
        transient brightenings and implications for the heating of the
        active-region corona}, PASJ 47, 251-263.}
\ref{Shukla, S., Brachet, M.E., and Pandit, R.
	2024, Phys.Rev. D, 110/8, id083002, 18pp.}
	{\sl Neutron-superfluid vortices and proton-superconductor 
	flux tubes: Development of a minimal model for pulsar glitches}
\ref{Silverberg, S.M., Kowalski, A.F., Davenport, R.A., Wisniewski, J.P.
	2016, ApJ 829:129 (11pp).}
	{\sl KELPLER flares IV. A comprehensive analysis of the activity
	of the dM4e star GJ 1243}
\ref{Souza, A.M., Echer, E., Bolzan, M.J.A., and Hajra, R.
	2016, J.Atmos.Solar-Terr.Phys. 149, p.81-86.}
	{\sl A study on the main periodicities in interplanetary magnetic
	field Bz component and geomagnetic AE index during HILDCAA
	events using wavelet analysis}
\ref{Spitler, L.G., Scholz, P., Hessels, J.W.T., Bogdanov, S. et al.
	2016, Nature, 531, Issue 7593, pp.202-205.}
	{\sl A repeating fast radio burst}
\ref{Svanda, M. and Karlicky, M.
	2016, ApJ 831, id 9, 7 pp.}
	{\sl flares on A-type stars: Evidence for heating of solar corona
	by nanoflares?}
\ref{Tavecchio, F., and Sobacchi.E.
	2020, MNRAS 491/2, p.2198-2204.}
	{\sl Anisotropic electron populations in BL Lac jets: 
	consequences for the observed emission}
\ref{Tavecchio, F., Bonnoli, G., and Galanti, G.
	2020a, MNRAS 497/1, pp.1294-1300.}
	On the distribution of fluxes of gamma-ray blazars: 
	hints for a stochastic process?
\ref{Tavecchio, F., Landoni, M., Siron, L., Coppi, P.
	2020b, MNRAS 498/1, pp.599-608.}
	{\sl Probing shock acceleration in BL Lac jets through 
	X-ray polarimetry: the time-dependent view}
\ref{Thibeault, C., Strugarek, A., Charbenneau, P., and Tremblay, B. 
	2022, SoPh 297, id.125.}
	{\sl Forecasting solar flares by data assimilation in sandpile
	models} 
\ref{Tu, Z.L., Yang, M., Zhang, Z.J. and Wang, F.Y.
	2020, ApJ 890:46 (15pp).}
	{\sl Superflares on solar-type stars from the first three
	year observation of TESS} 
\ref{Turolla, R., Zane, S., and Watts, A.L.
	2015, Rep.Prog.Phys. 78/11, id.116901.}
	{\sl Magnetars: the physics behind observations. A review}
\ref{Ulyanov, A.S., Bogachev, S.A., Reva, A.A., Kirichenko, A.S. et al.
	2019, Astron.Lett. 45/4, pp.248-257.}
	{\\sl The energy distribution of nanoflares at the minimum
	and rising phase of solar cycle}
\ref{Van Doorsselaere T., Shariari, Hoda, and Deboescher, J.
	2017, ApJSS 232, id.26, 13pp.}
	{\sl Stellar flares observed in long-cadence data from the
	KEPLER mission}
\ref{Verbeeck, C., Kraaikamp, E, Ryan, D.F., and Podladchikova, O. 
	2019, ApJ 884.50, (16pp).}  
	{\sl Solar flare distributions: Lognormal instead of power law?}
\ref{Verma, A.K., Pandit, R., and Brachet, M.E.
	2022, Phys.Rev. Research, 4/1, id.013026.}
	{\sl Rotating self-gravitating Bose-Einstein condensates 
	with a crust: A model for pulsar glitches}
\ref{Vilangot Nhalil Nived, Nelson, C.J., Mathioudakis, M. et al.
	2020, MNRAS 499 issue 1, pp.1385-1394.}
	{\sl Power-law energy distributions of small-scale impulsive events
	on the active Sun: Results from IRIS}
\ref{Vlahos, L. and Isliker, H.
	2016, European Phys. J. Special Topics 225/6, id.999pp.}
	{\sl Complexity methods applied to turbulence in plasma 
	astrophysics}
\ref{Wang, F.Y., Zhang, G.Q., and Dai, Z.G.
	2021, MNRAS 501/3, pp.3155-3161}
	{\sl Galactic and cosmological fast radio bursts as 
	scaled-up solar radio bursts} 
\ref{Wang, W.B., Li, C., Tu, Z.L., Guo, J.H. et al.
	2022, MNRAS 512, 1567-1673.}
	{\sl Avalanches of magnetic flux rope in the state of
	self-organized criticality}
\ref{Wang, Z.H., Sang, Y., and Zhang, X.
	2023a, RAA 23, id.025002, (5pp).}
	{\sl Power-law distribution and scale invariant
	structure from the First CHIME/FRB fast radio burst
	catalog}
\ref{Wang, Z.H., Sang, Y., and Z,X.
	2023b, ApJL 949/2, id.L33, 9pp.}
	{\sl Repeating Fast Radio Bursts Reveal Memory from Minutes to an Hour}
\ref{Wang, F.Y., Dai, Z.G., Yi, S.X. et al.
	2015, ApJSS 216/1, id.8, 8pp.}
	{\sl Universal behavior of X-ray flares from black hole systems}
\ref{Warszawski, L. and Melatos, A.
	2019, MNRAS 390/1, pp.175-191.}
	{\sl A cellular automaton model of pulsar glitches}
\ref{Watkins, N.W., Pruessner, G., Chapman, S.C., Crosby, N.B., et al. 
	2016, SSRv 198, 3.}
	{\sl 25 Years of Self-organized Criticality: Concepts and
        Controversies} 
\ref{Wei, J.J., Wu, X.F., Dai, Z.G., Wang, F.Y. et al.
	2021, ApJ 920:153 (7pp).}
	{\sl Similar scale-invariant behaviours between soft gamma-ray
	repeaters and an extreme epoch from FRB 121102}
\ref{Wei, J.J.
	2023, Phys.Rev Research, 5/1, id.013019.}
	{\sl Scale invariance in X-ray flares of gamma-ray bursts}
\ref{Wiegelmann, T. and Sakurai, T.
	2012, Living Rev. Sol. Phys. 9/1, id 5, 49pp.}
	{\sl Solar force-free magnetic fields}
\ref{Wu, C.J., Ip W.H., and Huang, L.C.
	2015, ApJ 798:92 (13pp).}
	{\sl A study of variability in the frequency distributions
	of the superflares of G-type stars observed by the KEPLER 
	mission}
\ref{Xiao, S., Yang, J.J., Luo, X.H., et al.
	2023a, ApJSS 268, id.5, 9 pp.}
	{\sl The minimum variation timescales of X-ray bursts
	from SGR J1935+2154}
\ref{Xiao, S., Tuo Y.L., Zhang, S.N. et al. 
	2023b, MNRAS 521/4, pp.5308-5333.}
	{\sl Discovery of the linear energy dependence of the 
	spectral lag of X-ray bursts from SGR J1935+2154}
\ref{Xiao, S., Zhang, Y.T., Yang, J.J. et al.
	2024a, ApJ 967, 128, 11 pp.}
	{\sl Log-Gaussian distribution and evolution of
	minimum variation timescales for SGR J1550+5418}
\ref{Xiao, S., Zhang, S.N., Xiong, S.L. et al.
	2024b, MNRAS 528/2, pp 1388-1392.
	{\sl The self-organized criticality behaviours of two new
	parameters in SGR J1935+2154.}
\ref{Xiao, S., Li, X.B., Xue, W.C., Yiong, S.L. et al.}
	2024c, MNRAS 527/4, pp.11915-11924.}
	{\sl  Individual and averaged power density spectra
	of X-ray bursts from SGR J1935+2154: quasi-periodic
	search and slopes}
\ref{Xiao, S., Hong,, M.X., Mei, X.Y., et al.
	2024d, ApJS 274, 14.}
	{\sl The self-organized criticality behaviours of pulses in
	magnetar bursts}
\ref{Xiao, S., Zhang, Y.Q., Zhu, Z.P., et al. 
	2024e, ApJ 970, 6.}
	{\sl The peculiar precursor of a gamma-ray burst from a
	binary merger involving a magnetar}
\ref{Xie, S.L., Yu, Y.W., Xiong, L. et al.
	2024, ApJ 967/2, id.108, 9pp.}
	{\sl Neutron-superfluid vortices and proton-superconductor 
	flux tubes: Development of a minimal model for pulsar glitches}
\ref{Yan, D., Yang, S., Zhang, P., Dai, B., et al.
	2018, ApJ 864:164 (16pp).}
	{\sl Statistical analysis on XMM-Newton X-ray flares 
	of Mrk 421: Distributions of peak flux and flaring 
	time duration}
\ref{Yang, H. and Liu, J.
	2019, ApJSS 241, id.29, 19pp.}
	{\sl The flare catalog and the flare activity in the 
	KEPLER mission}
\ref{Yi, S.X., Xi, S.Q., Fu, H., Wng, F.Y. et al.
	2016, ApJSS 224:20 (13pp).}
\ref{Yu, M. and Liu, Q.J.
        2017, MNRAS 468/3, p.3031-3041.}
        {\sl On the detection probability of neutron star glitches}
\ref{Yuan, Y.F.
        2017, Scientia Sinica Physica, Mech.Aster. 47/1, p.010402}
        {\sl Black hole binaries in the universe}
\ref{Yuan, Q., Wang, Q.D., Liu, S., Wu, K.
        2018, MNRAS 473/1, p.306-316.}
        {\sl A systematic Chandra study of Sagittarius $A^*$: 
	II. X-ray flare statistics}
\ref{Zhang, J., Liang, E.W., Zhang, S.N., and Bai, J.M.
	2012, ApJ 752/2, id.157, 18pp.}
	{\sl Radiation Mechanisms and Physical Properties of 
	GeV-TeV BL Lac Objects}
\ref{Zhang, H.M., Zhang, J., Lu, R.J. Yi, F.F. et al.
	2018, RAA 18/4, id.040.}
	{\sl Flux and spectral variation chaacteristics of
	3D 454.3 at the GeV band}
\ref{Zhang, G.Q., Wang, P., Wu, Q., Wang, F.Y. et al.
	2021, ApJ 920:L23 (11pp.)}
	{\sl Energy and waiting time distributions of FRB 121102
	observed by FAST}
\ref{Zhang, W.L., Yi, S.X., Yang, Y.P., Qin, Y.
	2022, RAA 22/6, id.065012, 8pp.}
	{\sl Statistical properties of X-ray flares from the
	supergiant fast X-ray transients} 
\ref{Zhang, W.L., Li, X.J., Yang, Y.P., et al.
	2023, RAA 23/11, id.115013, 6pp.}
	{\sl Statistical properties of X-ray bursts from
	SGR J1935+2154 by Insight-HXMT}
\ref{Zhang, W.L., Yi, S.X., Zou, Y.C., Wang, F.Y. et al. 
        2025, Astron Astrophys. 693, id.A290, 5pp.}
	{\sl Self-organized critical characteristics of
	Tera-electronvol photons from GRB 221009A}
	
\clearpage

\begin{figure}
\centerline{\includegraphics[width=0.9\textwidth]{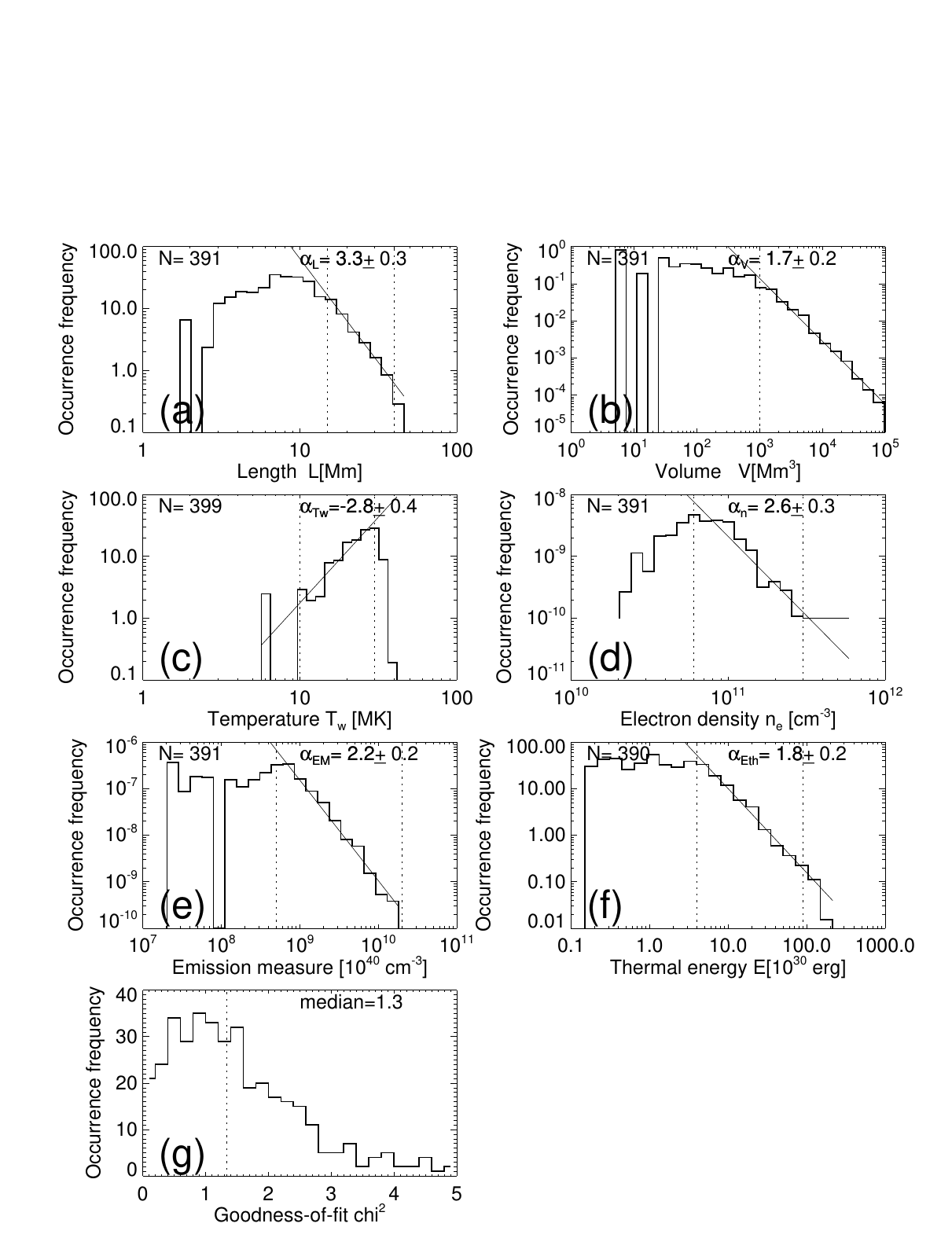}}
\caption{Size distributions of the physical parameters
$L, V, T_w, n_e, EM$ and $E_{th}$ for 391 analyzed M and X-class
flares. A powerlaw function is fitted in the range indicated with
dotted vertical lines. The reduced $\chi^2$ distribution is 
characterized with a median value of $\chi^2 = 1.3$ 
(Aschwanden et al.~2015a).}
\end{figure}

\begin{figure}
\centerline{\includegraphics[width=0.9\textwidth]{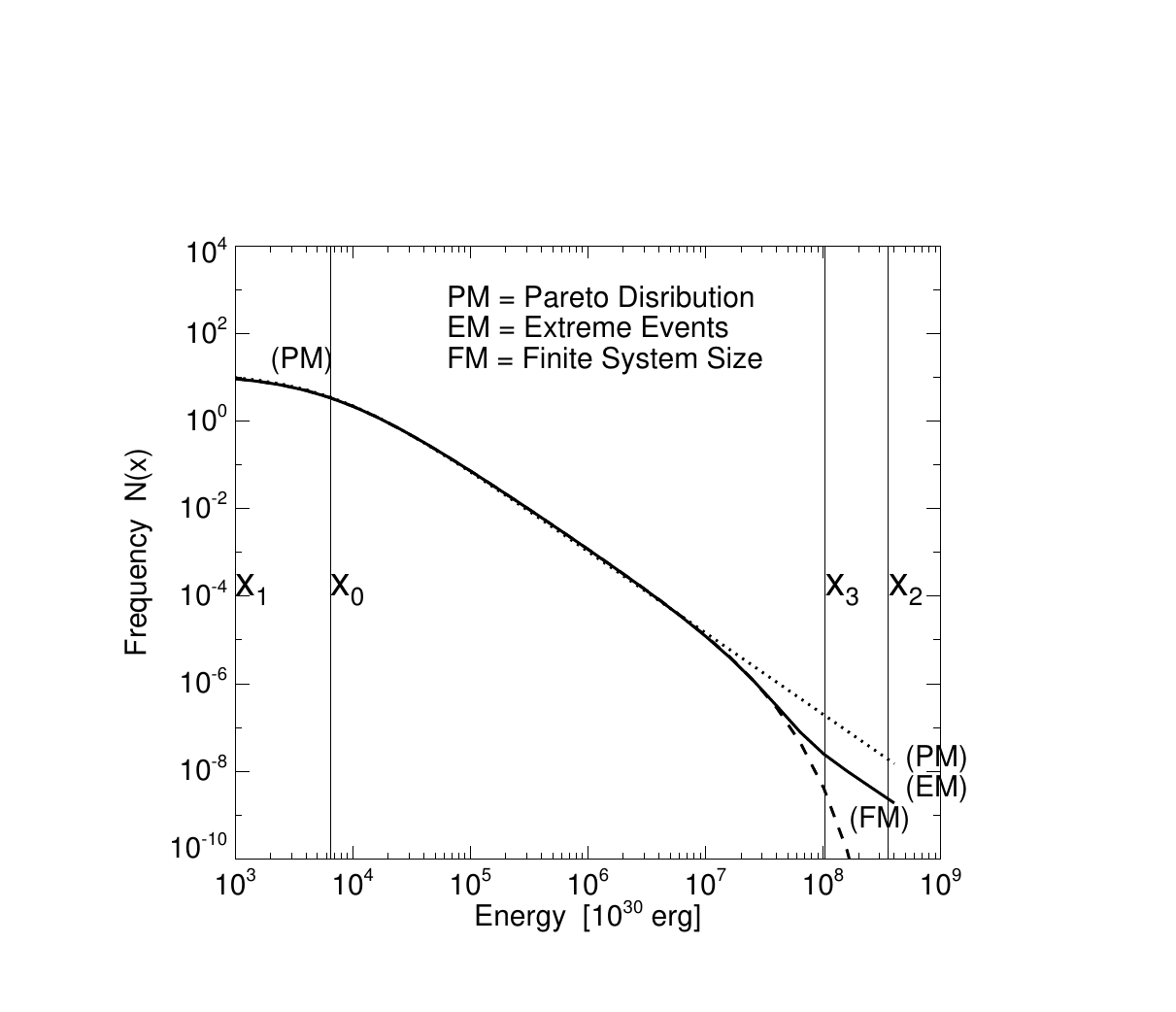}}
\caption{Synopsis of three power law models:
(PM) = Pareto distribution model $[x_1, x_0]$,
(EM) = Extreme events model $[x_3, x_2$], and
(FM) = Finite system-size model $[\gapprox x_3]$.
The inertial range covers $[\gapprox x_0, \lapprox x_3$].
The data used here are from a stellar flare catalog
observed with Kepler (Aschwanden 2021).}
\end{figure}

\begin{figure}
\centerline{\includegraphics[width=1.0\textwidth]{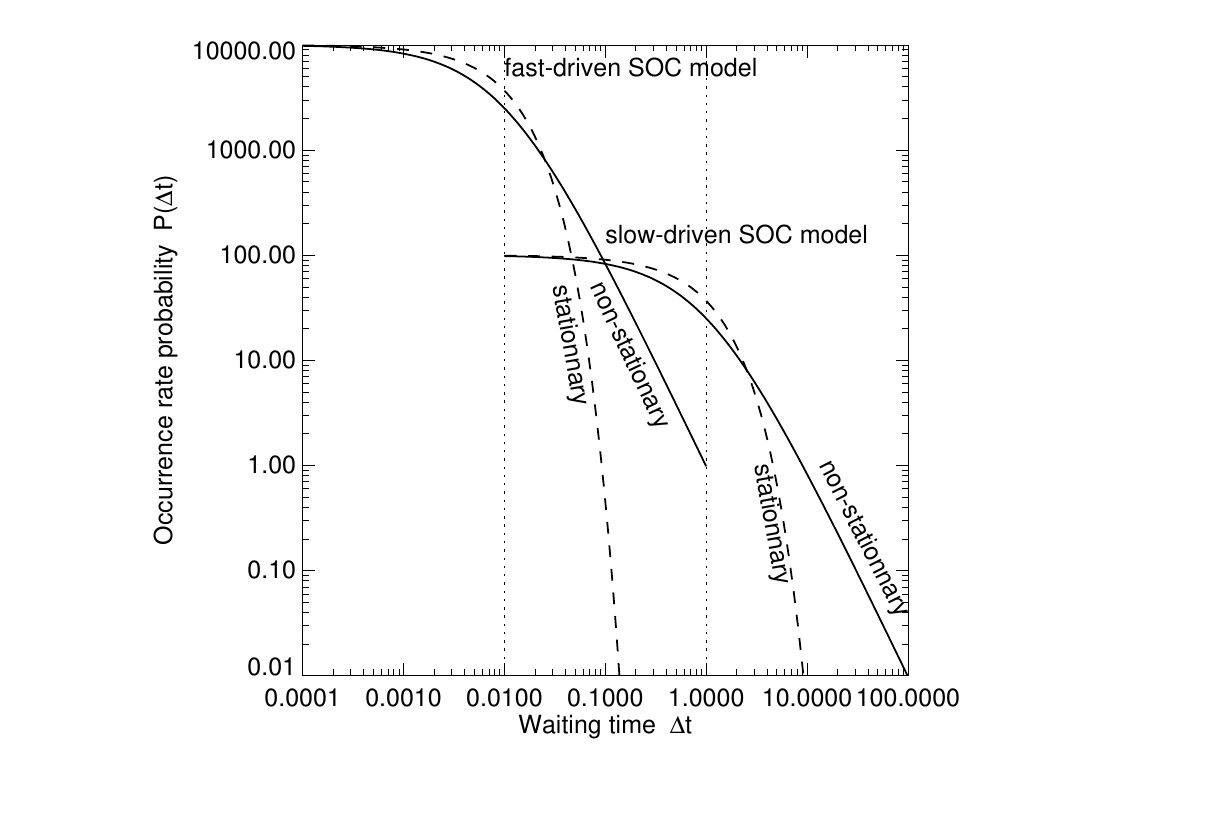}}
\caption{Overview of waiting time distributions for slow-driven
and fast-driven, stationary and non-stationary SOC models: 
Stationary models imply exponential WTD functions,
while nonstationary models produce power law size distributions
(Aschwanden 2019c). } 
\end{figure}

\begin{figure}
\centerline{\includegraphics[width=0.8\textwidth,angle=270]{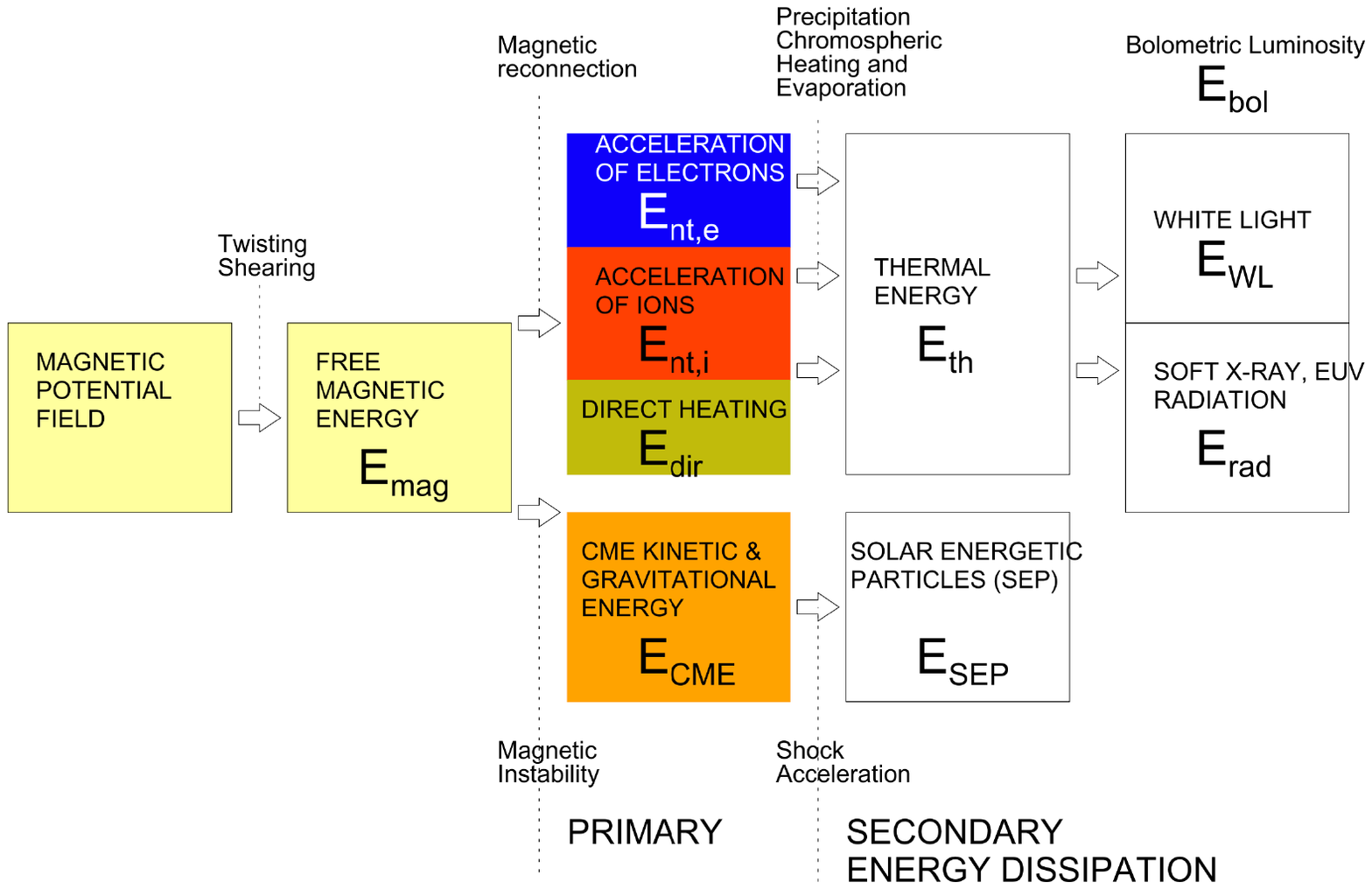}}
\caption{Schematic diagram of energy input (free magnetic 
energy $E_{\rm mag}$), primary energy dissipation processes 
(electron acceleration $E_{nt,e}$, ion acceleration $E_{nt,i}$,
direct heating $E_{dir}$, and launching of CME $E_{\rm CME}$), 
and secondary energy dissipation processes (thermal energy 
$E_{th}$, solar energetic particles $E_{\rm SEP}$, and bolometric
luminosity $E_{\rm bol}$, with radiative energies observed 
in white light $E_{\rm WL}$, and soft X-rays and EUV $E_{\rm rad}$),
(Aschwanden et al.~2017).}
\end{figure}

\begin{figure}
\centerline{\includegraphics[width=1.0\textwidth]{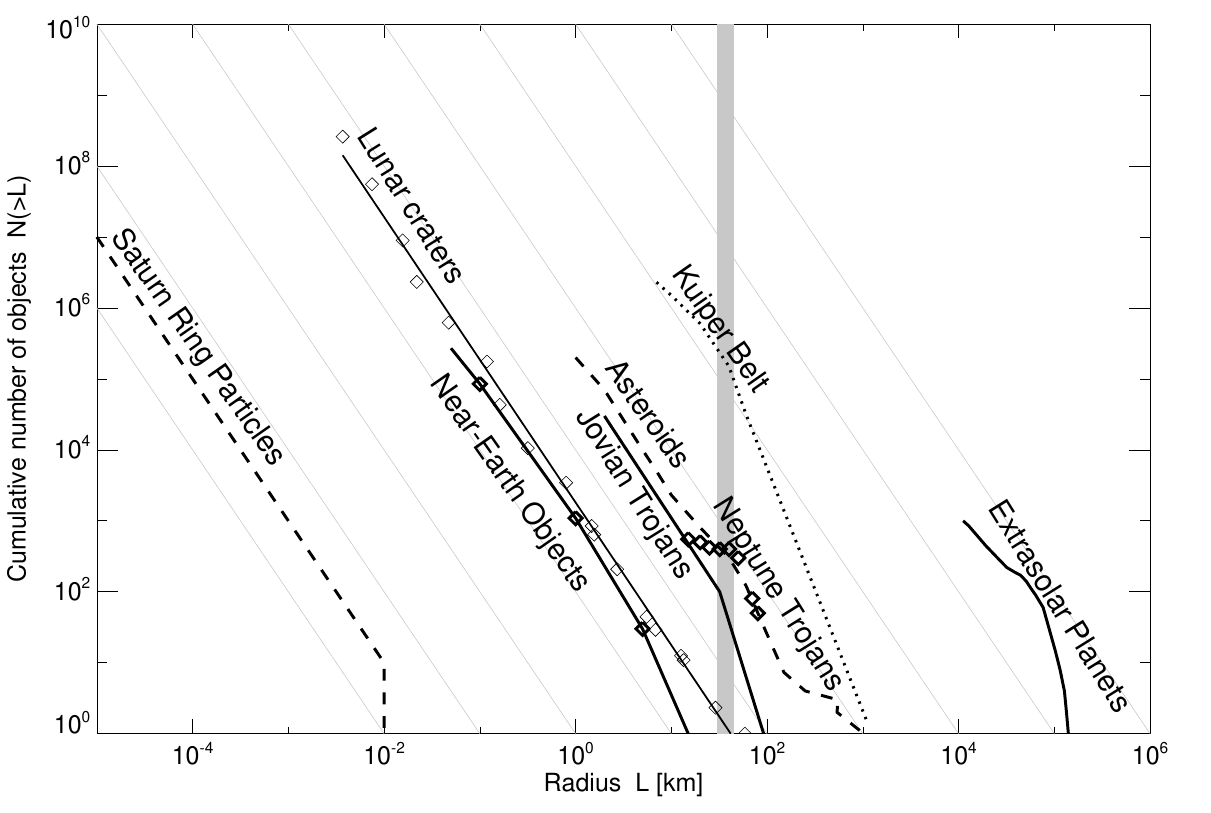}}
\captio{Cumulative size distribution of
Saturn ring particles, near-Earth objects, Jovian Trojans,
asteroids, Neptunian Trojans, lunar craters, Kuijper belt objects,
Neptune Trojans,and Earth-sized extra-solar planets.
The grey diagonal lines indicate the prediction of the FD-SOC model,
with a power law slope of $\alpha_L^{cum} = 2$ for the cumulative size
distribution, corresponding to a power law slope of $\alpha_L
=\alpha_L^{cum}+1=3$ for the differential occurrence frequency distribution.
References are given in Aschwanden et al.~(2016b).}
\end{figure}

\begin{figure}
\centerline{\includegraphics[width=1.0\textwidth]{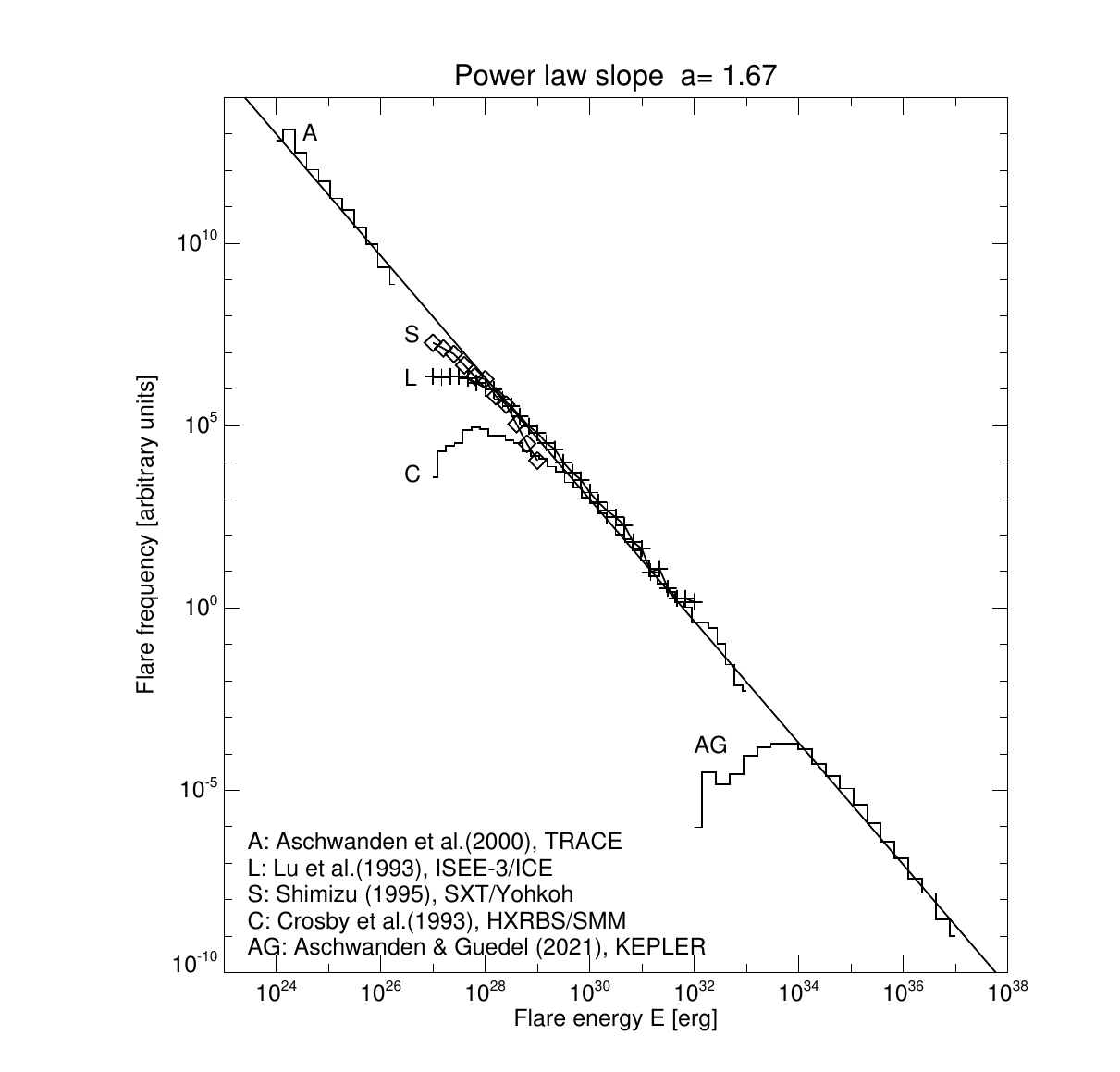}}   
\caption{Energy size distributions synthesized from solar
observations with HXRBS/SMM, ISEE-3/ICE, SXT/Yohkoh, TRACE   
and from stellar observaions with KEPLER. The theoretical
predictions of the power law slope for the flare energy
distribution is $\alpha_E = 1.67$ in the FD-SOC model.
The flare frequencies are aligned to the predicted power 
law size distributions (Aschwanden and Schrijver 2025).} 
\end{figure}

\end{document}